\documentclass[draft]{agujournal2019}
\usepackage{url} 
\usepackage[inline]{trackchanges} 
\usepackage{soul}
\usepackage{amsmath}
\usepackage{esint}
\draftfalse

\journalname{JGR: Space Physics}

\DeclareMathOperator*{\argmin}{arg\,min}

\begin{document}

\title{Time-domain modeling of 3-D Earth's and planetary electromagnetic induction effect in ground and satellite observations}

%
%

\authors{Alexander V. Grayver\affil{1}, Alexey Kuvshinov\affil{1}, Dieter Werthm\"uller\affil{2}}

\affiliation{1}{Institute of Geophysics, ETH Zurich, Sonneggstrasse 5, CH-8092 Zurich}
\affiliation{2}{Faculty of Civil Engineering and Geosciences, TU Delft, Netherlands}  

\correspondingauthor{Alexander V. Grayver}{agrayver@erdw.ethz.ch}

\begin{keypoints}
\item Accurate modeling of EM induction effects in ground and satellite observations via local and global impulse responses;
\item Including 3-D EM induction effects improves description of the observed magnetic variations during both quiet and disturbed conditions;
\item We provide a data constrained model of external currents for the \textit{Swarm} era;
\end{keypoints}

%
%

\begin{abstract}
Electric currents induced in conductive planetary interiors by time-varying magnetospheric and ionospheric current systems have a significant effect on electromagnetic (EM) field observations. Complete characterization of EM induction effects is difficult owing to non-linear interactions between the three-dimensional (3-D) electrical structure of a planet and spatial complexity of inducing current systems. We present a general framework for time-domain modeling of 3-D EM induction effects in heterogeneous conducting planets. Our approach does not assume that the magnetic field is potential, allows for an arbitrary distribution of electrical conductivity within a planet, and can deal with spatially complex time-varying current systems. The method is applicable to both data measured at stationary observation sites and satellite platforms, and enables the calculation of 3-D EM induction effects in near real-time settings.
\end{abstract}

%
%

\section{Introduction}

The effect of electric (telluric) currents induced in subsurface was observed in time series of geomagnetic field variations as early as in \citeA{schuster1889}, where it was also proposed that this effect depends on the electrical conductivity at depth. Subsequent studies have led to the establishment of an entirely new research field that exploits the electromagnetic (EM) induction phenomenon to sound planetary interiors \cite{price1967}. Nevertheless, present studies focusing on natural current systems, such as in magnetosphere or ionosphere, often neglect the effect of currents induced in the subsurface or treat it by using a variety of simplistic assumptions. However, as model parameterizations have become more realistic and accuracy of the geomagnetic measurements has improved, the effect of induction may no longer be neglected or substantially simplified, creating a need for efficient methods which can accurately account for it.

There exist two principal approaches to account for the induction effects in geomagnetic data. First, one can separate an observed vector magnetic field into inducing (external) and induced (internal) parts by using the classic Gauss' method \cite{gauss1877}. However, limitations imposed by this method, namely that the magnetic field must be potential and measured in a region between the inducing and induced currents, either restrict or invalidate its applicability. Besides, including more unknowns in statistical models to constrain the induced part may quickly degrade the quality of the models given noisy data with limited coverage. An alternative to that is to model the EM induction effect due to extraneous currents by invoking the governing Maxwell's equations. 

The latter approach has several advantages. Unlike Gauss method, the modeling approach is applicable regardless of the position of measurements relative to the inducing and induced current regions, and remains valid in regions where the field is not potential. Additionally, this has a positive effect for the conditioning of statistical models since extra unknowns used to describe the induced part can be eliminated. 

In practice, the complication behind modeling EM induction in geomagnetic observations is twofold. First, one needs to assume a subsurface conductivity model. A number of regional and global conductivity models exist. This study will not focus on how these models are constructed and whether they represent the subsurface accurately, even though inaccurate conductivity models may bias results. It is expected that our knowledge about the electrical structure of the subsurface will continuously improve, allowing for the construction of more accurate models at different scales \cite{kelbert2020}. Second, even if a distribution of the subsurface conductivity was known, modeling induced response of a 3-D heterogeneous planet remains a computationally demanding problem. 

Our goal is to develop an efficient time-domain method for calculating the EM induction effect of a planet with an arbitrary 3-D conductivity distribution that is suitable for both ground and satellite observations. 

One way to calculate a planet's EM induction effect is through frequency domain (FD) transfer functions, which describe a planet's response due to ''elementary'' extraneous currents. Modeling 3-D EM induction effects with transfer functions was previously applied to analyse daily magnetic field variations \cite{yamazaki2017sq} in ground \cite{kuvshinov1999global, koch2013global, guzavina2019probing} and satellite measurements \cite{sabaka2004extending, sabaka2015, chulliat2016, sabaka2018comprehensive}. Additionally, it was applied in the analysis of aperiodic geomagnetic variations in ground observations \cite{olsen2004modeling, puthe2014reproducing, sun2015ionospheric, honkonen2018predicting, munch2020joint}. These studies Fourier transformed data and applied transfer functions in frequency domain, followed by inverse Fourier transform in order to obtain results in time domain. Therefore, all aforementioned studies effectively worked in frequency domain.

However, the FD approach based on transfer functions has limitations in many practical scenarios. Among them are applications involving nearly real-time predictions of induction effects with constantly augmented time series, such as space weather hazard assessment, or estimation of steering errors in geomagnetic navigation while drilling. The limitations of the FD approach are also apparent when working with data from constantly moving satellites due to spatio-temporal aliasing. To overcome these restrictions, transfer functions can be converted into impulse responses and applied to the data directly in time domain. This approach was adopted by \citeA{maus2004, olsen2005, thomson2007} for modeling EM induction effects in satellite data. However, these works only considered the induction effect due to an external source described by a single (first zonal) spherical harmonic function and, moreover, assuming a 1-D subsurface conductivity distribution. The extension of this concept to general settings and presentation of all methodological details constitute the main contribution of this study.

Here, we calculate time-domain impulse responses of a medium by converting transfer functions pre-calculated in frequency domain. We achieve high computational efficiency by applying optimal digital linear filters (DLF) \cite{ghosh1970phd, ghosh1971dlfa} with the lagged convolution method \cite{anderson1975improved}, which require only a small set of (computationally expensive) frequency domain solutions. For this purpose, we design new DLFs using the methodology presented in \citeA{werthmuller2019tool}. Alternatively, evaluation of impulse responses of a 3-D medium can be done by means of dedicated time-domain induction solvers \cite{velimsky2003transient, velimsky2005time}. 

The methods developed here are applied to describe induction effect due to ionospheric and magnetospheric currents in ground and satellite geomagnetic observations. However, the formalism is amenable to observations made around other planets, where conventional methods may be too restrictive \cite<e.g.>[]{olsen2010}.

\section{Methods}

\subsection{Governing equations}

Electromagnetic field variations are governed by Maxwell's equations. In frequency domain, these equations read
\begin{eqnarray}
\frac{1}{\mu_0} \nabla \times \vec{B} &=& \sigma \vec{E} + \vec{j}^{\textrm{ext}}, \\
\nabla \times \vec{E} &=& - \textrm{i} \omega \vec{B}, 
\label{eq:maxwell}
\end{eqnarray}
where $\mu_0$ is the magnetic permeability of free space; $\omega$ angular frequency; $\vec{j}^{\textrm{ext}}(\vec{r},\omega)$ the extraneous (impressed) electric current density; $\vec{B}(\vec{r}, \omega; \sigma), \vec{E}(\vec{r}, \omega; \sigma)$ are magnetic and electric fields, respectively; $\sigma(\vec{r})$ spatial distribution of electrical conductivity; vector $\vec{r}=(r, \vartheta, \varphi)$ describes a position in the spherical coordinate system with $r$, $\vartheta$ and $\varphi$ being distance from the planet's centre, co-latitude, and longitude, respectively. Note that we neglected displacement currents and adopted the following Fourier convention
\begin{equation}
f(t)=\frac{1}{2\pi}\int\limits_{-\infty}^{\infty}\tilde{f}(\omega)e^{\mathrm{i}\omega t}\mathrm{d}\omega.
\label{eq:fourier}
\end{equation}

We assume that the current density, $\vec{j}^{\textrm{ext}}(\vec{r}, \omega)$, can be represented as a linear combination of spatial modes $\vec{j}_i(\vec{r})$, 
\begin{eqnarray}
\label{eq:ext_lincomb}
\vec{j}^{\textrm{ext}}(\vec{r}, \omega) &=& \sum_i  \vec{j}_i(\vec{r}) c_i(\omega),
\end{eqnarray}
where $\vec{j}_i(\vec{r})$ can, in practice, include electric dipoles, current loops \cite{sun2012} or be a continuous function. 

By virtue of the linearity of Maxwell's equations with respect to the $\vec{j}^{\textrm{ext}}(\vec{r}, \omega)$ term, we can expand total (i.e. inducing  plus induced) EM field as a linear combination of individual fields $\vec{B}_i, \vec{E}_i$,
\begin{eqnarray}
\vec{B}(\vec{r}, \omega; \sigma) &=& \sum_i \vec{B}_i(\vec{r}, \omega; \sigma) c_i(\omega) \label{eq:b_field_sum}, \\
\vec{E}(\vec{r}, \omega; \sigma) &=& \sum_i \vec{E}_i(\vec{r}, \omega; \sigma) c_i(\omega) \label{eq:e_field_sum}.
\end{eqnarray}

The $\vec{B}_i(\vec{r},\omega; \sigma)$ and $\vec{E}_i(\vec{r},\omega; \sigma)$ fields are solutions of the equations
\begin{eqnarray}
\frac{1}{\mu_0} \nabla \times \vec{B}_i &=& \sigma \vec{E}_i + \vec{j}_i \label{eq:maxwell_mode_1},\\
\nabla \times \vec{E}_i &=& - \textrm{i} \omega \vec{B}_i ,
\label{eq:maxwell_mode_2}
\end{eqnarray}
and, following definitions in \ref{sec:imp_resp}, represent EM transfer functions of a medium.

Therefore, a transfer function of a planet at a position $\vec{r}$ depends on the subsurface conductivity distribution and frequency of excitation as well as on the spatial geometry of the current density expressed through the $\vec{j}_i$ term. 

\subsection{Current density representation}

We now elaborate on the form of the current density term $\vec{j}^{\textrm{ext}}$. In this study, we assume that electric currents flow within an insulated spherical shell above the ground. This allows us to collapse any current density distribution within the shell into a current sheet characterized by a stream function
\begin{equation}
\label{eq:ext_stream}
\vec{j}^{\textrm{ext}}(\vec{r}, \omega) = -\delta(r-b)\hat{e}_r \times \nabla_H \Psi(\theta, \phi, \omega),
\end{equation}
where $a$ is planet's radius, $b = a + h$, with $h$ being the altitude of the current sheet,
\begin{equation}
\label{eq:nabla_H}
\nabla_H f = \frac{1}{r}\frac{\partial f}{\partial\theta}\hat{e}_{\theta} + \frac{1}{r\sin{\theta}}\frac{\partial f}{\partial\phi}\hat{e}_{\phi},
\end{equation}
and $\hat{e}_r$, $\hat{e}_{\theta}$ and $\hat{e}_{\phi}$ are the unit vectors of the spherical coordinate system. Consequently, we can expand the stream function as a linear combination of spatial modes and scalar coefficients, that is
\begin{equation}
\label{eq:lin_comb_phi}
\Psi(\theta, \phi, \omega) = \sum_i \Psi_i(\theta, \phi) c_i(\omega).
\end{equation}
Using eqs. (\ref{eq:ext_lincomb}) and (\ref{eq:lin_comb_phi}), we can rewrite eq. (\ref{eq:ext_stream}) as
\begin{eqnarray}
\label{eq:ext_lincomb_phi}
\vec{j}^{\textrm{ext}}(\vec{r}, \omega) = -\delta(r-b) \sum_i \left[ \hat{e}_r \times \nabla_H \Psi_i(\theta, \phi)\right]c_i(\omega).
\end{eqnarray}

\subsection{Spherical harmonic representation}

The choice of spatial functions $\Psi_i$ is generally problem dependent. In this study, we will adopt spherical harmonic (SH) representation. Then, for an external source, a stream function can be written as \cite{schmucker1985}
\begin{equation}
\label{eq:stream_sh}
\Psi^e(\vec{r}, \omega) = -\frac{a}{\mu_0}\sum_{(n, m) \in \mathcal{M}} \frac{2n+1}{n+1}\left(\frac{b}{a}\right)^n \tilde{\varepsilon}_n^m(\omega) S_n^m(\theta, \phi),
\end{equation}
where 
\begin{equation}
S_n^m(\theta, \phi) = P_n^{|m|}(\cos{\theta})\exp{(\textnormal{i}m\phi)}
\end{equation}
is a spherical harmonic (SH) function of degree $n$ and order $m$ with $P_n^{|m|}$ being Schmidt semi-normalized associated Legendre polynomials and $\mathcal{M}$ is a set of SH functions with corresponding complex-valued SH coefficients $\tilde{\varepsilon}_n^m(\omega)$. 

This allows us to rewrite eq. (\ref{eq:ext_lincomb_phi}) as
\begin{eqnarray}
\vec{j}^{\textrm{ext}}(\vec{r}, \omega) &=& \sum_{(n, m) \in \mathcal{M}} \vec{j}_n^m(\vec{r}) \tilde{\varepsilon}_n^m(\omega),
\end{eqnarray}
with
\begin{equation}
\label{eq:jnm}
\vec{j}_n^m(\vec{r}) = \frac{\delta(r-b)}{\mu_0} \frac{2n+1}{n+1}\left(\frac{b}{a}\right)^{n-1} \hat{e}_r \times \nabla_{\perp} S_n^m(\theta, \phi),
\end{equation}
where $\nabla_{\perp} = r\nabla_H$. Accordingly, following eqs. (\ref{eq:b_field_sum})-(\ref{eq:e_field_sum}), total electric and magnetic fields at a position $\vec{r}$ can be expressed as 
\begin{eqnarray}
\label{eq:unit_field_sh_b}
\vec{B}(\vec{r}, \omega; \sigma) &=& \sum_{(n, m) \in \mathcal{M}} \vec{B}_n^m(\vec{r}, \omega; \sigma) \tilde{\varepsilon}_n^m(\omega), \\
\label{eq:unit_field_sh_e}
\vec{E}(\vec{r}, \omega; \sigma) &=& \sum_{(n, m) \in \mathcal{M}} \vec{E}_n^m(\vec{r}, \omega; \sigma) \tilde{\varepsilon}_n^m(\omega),
\end{eqnarray}
where $\vec{B}_n^m, \vec{E}_n^m$ are magnetic and electric field transfer functions due to the current density distribution as given by eq. (\ref{eq:jnm}). In what follows, we will work with the magnetic field only, although some applications in the field of space weather modeling may take advantage of eq. (\ref{eq:unit_field_sh_e}) to work with electric fields.

Note that eqs. (\ref{eq:stream_sh})-(\ref{eq:unit_field_sh_e}) are only valid for a source that is external relative to the observer. The equivalent derivations for internal sources (such as, for example, ionosphere in satellite data) can be carried out by taking \cite{schmucker1985}
\begin{equation}
\label{eq:stream_sh_int}
\Psi^i(\vec{r}, \omega) = \frac{a}{\mu_0}\sum_{(n, m) \in \mathcal{M}} \frac{2n+1}{n}\left(\frac{a}{b}\right)^{n+1} \tilde{\iota}_n^m(\omega) S_n^m(\theta, \phi)
\end{equation}
instead of eq. (\ref{eq:stream_sh}).

\subsection{Impulse responses and transfer functions}

In this section, we present methods to calculate EM signals induced by an electric current of the form (\ref{eq:jnm}) and measured on the ground or in space. 

\subsubsection{Local impulse responses}
\label{sec:local}

For reasons that we discussed in the introduction, it is often more convenient to work with data in time domain. Therefore, total magnetic field at a location $\vec{r}$ and time $t$ can be best described by eq. (\ref{eq:unit_field_sh_b}) after its transformation to time domain. Eq. (\ref{eq:unit_field_sh_b}) can be written in time domain as a convolution integral (see \ref{sec:imp_resp} for more details)
\begin{equation}
\label{eq:field_via_kernels}
\vec{B}(\vec{r}, t; \sigma) = \sum_{(n, m) \in \mathcal{M}^+}  \int_{-\infty}^{t} \left[ \vec{B}_n^{m(c)}(\vec{r}, t - \tau; \sigma)q_n^m(\tau) + \vec{B}_n^{m(s)}(\vec{r}, t - \tau; \sigma)s_n^m(\tau) \right] \textnormal{d}\tau,
\end{equation}
where $\mathcal{M}^+$ is a set of SH functions with non-negative orders ($m \geq 0$); $q, s$ inducing SH coefficients; $\vec{B}_n^{m(c)}$ and $\vec{B}_n^{m(s)}$ are impulse responses of a medium for the $q_n^m$ and $s_n^m$ coefficients, respectively. They can be defined as
\begin{equation}
\label{eq:bsin_cos}
\vec{B}_n^{m(c)}(\vec{r}, t; \sigma) = -\frac{2}{\pi}\int_{0}^{\infty}\mathrm{Im}\left[\frac{\vec{B}_n^m(\vec{r}, \omega; \sigma) + \vec{B}_n^{-m}(\vec{r}, \omega; \sigma)}{2}\right]\sin{(\omega t)}\mathnormal{d}\omega
\end{equation}
and
\begin{equation}
\label{eq:bsin_sin}
\vec{B}_n^{m(s)}(\vec{r}, t; \sigma) = \frac{2}{\pi}\int_{0}^{\infty}\mathrm{Im}\left[\frac{\vec{B}_n^m(\vec{r}, \omega; \sigma) - \vec{B}_n^{-m}(\vec{r}, \omega; \sigma)}{2\mathrm{i}}\right]\sin{(\omega t)}\mathnormal{d}\omega.
\end{equation}
The integrals in eqs. (\ref{eq:bsin_cos})-(\ref{eq:bsin_sin}) are evaluated by using the digital linear filter method as explained in \ref{sec:dlf}.

\subsubsection{Global impulse responses}
\label{sec:global}

For satellite measurements, using local impulse responses becomes impractical since it requires calculating eqs. (\ref{eq:bsin_cos})-(\ref{eq:bsin_sin}) for every satellite location. Therefore, to describe EM induction effects in satellite data, we resort to different transfer functions, namely $Q$-responses and $Q$-matrices, which enable factorization of spatial and temporal effects. We note, however, that while transfer functions in eq. (\ref{eq:unit_field_sh_b}) are valid everywhere, $Q$-responses and $Q$-matrices are valid only in regions where the magnetic field is potential.

Recall that if a magnetic field at a position $\vec{r}$ and time $t$ is potential, we have 
\begin{eqnarray}
\label{eq:potfield}
\vec{B}(\vec{r}, t; \sigma) = -\nabla \left[V^e(\vec{r}, t) + V^i(\vec{r}, t; \sigma)\right],
\end{eqnarray}
where inducing and induced parts of the potential are given by
\begin{eqnarray}
\label{eq:magpotential_ext}
  V^e(\vec{r}, t) &=& a\sum_{n = 1}^{N}\sum_{m = 0}^n \left[ q_n^m(t)\cos(m\phi) + s_n^m(t)\sin(m\phi) \right]\left( \frac{r}{a} \right)^n P_n^m(\cos{\theta}) \nonumber \\ 
  &=& \textnormal{Re}\left\{ a \sum_{n=1}^{N}\sum_{m = -n}^{n} \varepsilon_n^m(t) \left( \frac{r}{a} \right)^n S_n^m(\theta, \phi) \right\}
\end{eqnarray}
and
\begin{eqnarray}
\label{eq:magpotential_int}
  V^i(\vec{r}, t; \sigma) &=& a\sum_{k = 1}^{K}\sum_{l = 0}^k \left[ g_k^l(t; \sigma)\cos(l\phi) + h_k^l(t; \sigma)\sin(l\phi) \right]\left( \frac{a}{r} \right)^{k+1} P_k^l(\cos{\theta}) \nonumber \\ 
  &=& \textnormal{Re} \left\{ a \sum_{k=1}^{K}\sum_{l = -k}^{k} \iota_k^l(t; \sigma) \left( \frac{a}{r} \right)^{k+1} S_k^l(\theta, \phi) \right\},
\end{eqnarray}
where $N, K$ are some constants that truncate series.
Note that we stated the magnetic field potential using both real-valued and complex-valued notations with the following relation between the coefficients,
\begin{equation}
\label{eq:complex_real_sh}
\varepsilon_n^m = 
\begin{cases}
\frac{q_n^m - \textnormal{i}s_n^m}{2}, & m > 0 \\
\frac{q_n^{|m|} + \textnormal{i}s_n^{|m|}}{2}, & m < 0 \\
q_n^m, & m = 0
\end{cases}.
\end{equation}
The relation between induced (internal in our case) coefficients $g_k^l, h_k^l$ and $\iota_k^l$ is derived in an identical way. 

We can now rewrite the induced magnetic field (\ref{eq:magpotential_int}) using transfer functions instead of induced SH coefficients. Before presenting the general case, we first consider a case when planet's conductivity distribution is assumed to be 1-D, i.e., $\sigma(\vec{r}) \equiv \sigma(r)$. In this case, each coefficient $\varepsilon_n^m$ induces one internal coefficient of the same degree and order \cite<e.g.>[]{price1967}. Inducing and induced coefficients can be related via a scalar transfer function called $Q_n$-response. In frequency domain, this relation reads
\begin{equation}
\label{eq:qresponse}
\tilde{\iota}_n^m(\omega; \sigma) = \tilde{Q}_n(\omega; \sigma)\tilde{\varepsilon}_n^m(\omega).
\end{equation}
Note that $Q_n$ is independent of order $m$ \cite{schmucker1985}. 

Following derivations in \ref{sec:imp_resp}, transforming eq. (\ref{eq:qresponse}) to time domain and separating spatial sine and cosine terms leads to a pair of convolution integrals
\begin{eqnarray}
g_n^m(t; \sigma) &=& Q_n \ast q_n^m = \int_{-\infty}^t Q_n(t - \tau; \sigma)q_n^m(\tau) \textnormal{d}\tau \label{eq:qg_td}, \\
h_n^m(t; \sigma) &=& Q_n \ast s_n^m = \int_{-\infty}^t Q_n(t - \tau; \sigma)s_n^m(\tau) \textnormal{d}\tau \label{eq:qs_td}.
\end{eqnarray} 

Subsequently, substituting eqs. (\ref{eq:qg_td})-(\ref{eq:qs_td}) in eq. (\ref{eq:magpotential_int})  yields internal magnetic potential
\begin{equation}
\label{eq:magpotential}
  V^i(\vec{r}, t; \sigma) = V^{i(c)}(\vec{r}, t; \sigma) + V^{i(s)}(\vec{r}, t; \sigma)
\end{equation}
with
\begin{eqnarray}  
\label{eq:vint1d_cos}
   V^{i(c)}(\vec{r}, t; \sigma) &=& a\sum_{(n, m) \in \mathcal{M}^+} \left[Q_n \ast q_n^m\right] \cos({m\phi}) \left( \frac{a}{r} \right)^{n+1} P_n^m(\cos{\theta}), \\ 
\label{eq:vint1d_sin}
  V^{i(s)}(\vec{r}, t; \sigma) &=& a\sum_{(n, m) \in \mathcal{M}^+} \left[Q_n \ast s_n^m\right] \sin({m\phi}) \left( \frac{a}{r} \right)^{n+1} P_n^m(\cos{\theta}).
\end{eqnarray}
Note that in a 1-D case, inducing and induced expansions are identical, hence we used $n,m$ in eqs. (\ref{eq:qresponse})-(\ref{eq:vint1d_sin}) for all SH coefficients.

For a general 3-D conductivity distribution, $\sigma(\vec{r})$ in a planet, each coefficient $\varepsilon_n^m$ induces infinitely many internal coefficients \cite{olsen1999}. The relation between inducing and induced coefficients is then described by a set of transfer functions called $Q$-matrix
\begin{equation}
\label{eq:qmatrix}
\tilde{\iota}_k^l(\omega; \sigma) = \sum\limits_{n,m}\tilde{Q}_{kn}^{lm}(\omega; \sigma)\tilde{\varepsilon}_n^m(\omega).
\end{equation}
An element of the $Q$-matrix is given by \cite{Puethe2014_Q}
\begin{equation}
\label{eq:qnmkl}
\tilde{Q}_{kn}^{lm}(\omega; \sigma) = \frac{1}{(k + 1)\| S_k^l \|^2}\oiint_{\mathcal{S}(1)} \left[ B_{n, r}^m (\vec{r}_a, \omega; \sigma) - B_{n,r}^{m,\mathrm{ext}}(\vec{r}_a) \right]S_k^{l*}(\theta, \phi)\sin(\theta)\textnormal{d}\theta\textnormal{d}\phi,
\end{equation}
where $^*$ denotes complex conjugation, $\vec{r}_a = (a, \theta, \phi)$ is the position vector at the surface of a planet, and $\mathcal{S}(1)$ the surface of a ball with unit radius. The radial magnetic field $B_{n, r}^m$ is (numerically) computed for a given 3-D Earth's model induced by a unit amplitude ($\tilde{\varepsilon}_n^m$ = 1) SH current source described by eq. (\ref{eq:jnm}), and
\begin{equation}
B_{n,r}^{m,\mathrm{ext}}(\vec{r}_a) = -n S_n^m(\theta, \phi)
\end{equation}
is the inducing (external) part of the radial magnetic field. 

In this case, the internal magnetic potential becomes
\begin{eqnarray}  
\label{eq:vint3d_cos}
   V^{i(c)}(\vec{r}, t; \sigma) &=& a\sum_{(n, m) \in \mathcal{M}^+}\sum_{k,l} \left[Q_{kn}^{lm,qg} \ast q_n^m + Q_{kn}^{lm,sg} \ast s_n^m\right] \cos({k\phi}) \left( \frac{a}{r} \right)^{k+1} P_k^l(\cos{\theta}), \nonumber\\~ \\
\label{eq:vint3d_sin}
  V^{i(s)}(\vec{r}, t; \sigma) &=& a\sum_{(n, m) \in \mathcal{M}^+}\sum_{k,l} \left[Q_{kn}^{lm,qh} \ast q_n^m + Q_{kn}^{lm,sh} \ast s_n^m\right] \sin({k\phi}) \left( \frac{a}{r} \right)^{k+1} P_k^l(\cos{\theta}), \nonumber \\
\end{eqnarray}
where
\begin{equation}
\sum\limits_{k,l}=\sum\limits_{k=1}^{K}\sum\limits_{l=0}^{k}.
\label{eq:doublesums}
\end{equation}

After some algebra, impulse responses in eqs. (\ref{eq:vint3d_cos})-(\ref{eq:vint3d_sin}) can be calculated via sine transform (see eq. \ref{eq:sin_transform}) of the spectra, which are related to the frequency domain $Q$-matrix (eq. \ref{eq:qmatrix}) via equations below (the dependence on $\omega$ and $\sigma$ is omitted). 

\noindent For $l > 0, m > 0$:
\begin{eqnarray}  
\tilde{Q}_{kn}^{lm,qg} &=& \frac{\tilde{Q}_{kn}^{lm}+\tilde{Q}_{kn}^{l-m}+\tilde{Q}_{kn}^{-lm}+\tilde{Q}_{kn}^{-l-m}}{2},\\
\tilde{Q}_{kn}^{lm,qh} &=& \textnormal{i}\frac{\tilde{Q}_{kn}^{lm}+\tilde{Q}_{kn}^{l-m}-\tilde{Q}_{kn}^{-lm}-\tilde{Q}_{kn}^{-l-m}}{2},\\
\tilde{Q}_{kn}^{lm,sg} &=& \textnormal{i}\frac{-\tilde{Q}_{kn}^{lm}+\tilde{Q}_{kn}^{l-m}-\tilde{Q}_{kn}^{-lm}+\tilde{Q}_{kn}^{-l-m}}{2},\\
\tilde{Q}_{kn}^{lm,sh} &=& \frac{\tilde{Q}_{kn}^{lm}-\tilde{Q}_{kn}^{l-m}-\tilde{Q}_{kn}^{-lm}+\tilde{Q}_{kn}^{-l-m}}{2},
\end{eqnarray}
for $l = 0, m > 0$:
\begin{eqnarray}  
\tilde{Q}_{kn}^{0m,qg} &=& \frac{\tilde{Q}_{kn}^{0m}+\tilde{Q}_{kn}^{0-m}}{2},\\
\tilde{Q}_{kn}^{0m,sg} &=& \textnormal{i}\frac{-\tilde{Q}_{kn}^{0m}+\tilde{Q}_{kn}^{0-m}}{2},
\end{eqnarray}
for $l > 0, m = 0$:
\begin{eqnarray}  
\tilde{Q}_{kn}^{l0,qg} &=& \tilde{Q}_{kn}^{l0}+\tilde{Q}_{kn}^{-l0},\\
\tilde{Q}_{kn}^{l0,qh} &=& \textnormal{i}(\tilde{Q}_{kn}^{l0}-\tilde{Q}_{kn}^{-l0}),
\end{eqnarray}
and for $l = 0, m = 0$:
\begin{eqnarray}  
\tilde{Q}_{kn}^{00,qg} &=& \tilde{Q}_{kn}^{00}.
\end{eqnarray}

Note that both internal potentials, eqs. (\ref{eq:vint1d_cos})-(\ref{eq:vint1d_sin}) and eqs. (\ref{eq:vint3d_cos})-(\ref{eq:vint3d_sin}), depend only on the pre-calculated $Q$ and inducing coefficients. Additionally, $Q$ does not depend on location, making it particularly well-suited for satellite data. 

\subsection{Determination of inducing coefficients}
\label{sec:ext_coeffs}

The methods presented in the previous sections enable estimation of time-series of inducing coefficients in discrete non-overlapping time intervals (time windows). Let us define time intervals of length $\Delta t$. We assume that inducing coefficients are piece-wise constant within these time intervals. Then, convolution integrals such as eq. (\ref{eq:field_via_kernels}) or eqs. (\ref{eq:qg_td})-(\ref{eq:qs_td}) can be approximated by discrete sums. For instance, for a time window centered at $t$ we can rewrite eq. (\ref{eq:qg_td}) as
\begin{equation}
g_n^m(t; \sigma) \approx \sum_{j = 0}^{N_t} I_{Q_n}(j; \sigma)q_n^m(t - j\Delta t),
\end{equation}
where
\begin{equation}
\label{eq:qn_discrete}
I_{Q_n}(j; \sigma) = \int_{j\Delta t -\Delta t/2}^{j\Delta t +\Delta t/2} Q_n(t; \sigma)\mathrm{d}t.
\end{equation}
Similar expressions are obtained for other convolution integrals. 

With this, coefficients for a time window centered at $t$ can be estimated by solving a minimization problem,
\begin{equation}
\label{eq:minprob}
\mathbf{q}^{\ast}, \mathbf{s}^{\ast} = \argmin_{\mathbf{q}, \mathbf{s}} \sum_{i \in \mathcal{D}_t} \sum_{\alpha \in \{ \theta, \phi \}} \left[ B_{\alpha, i}^o - \sum_{(n, m) \in \mathcal{M}^+} B_{n,\alpha}^m(\vec{r}_i, t; \sigma) \right]^2, 
\end{equation}
where $\mathcal{D}_t$ is a set of magnetic field observations in the current time window with $B_{\alpha, i}^o$  being the measured  horizontal magnetic field component at location $\vec{r}_i$ and time $t_i$; $\mathbf{q}, \mathbf{s} \in \mathcal{M}^+$ are vectors of inducing SH coefficients for the given time window; and the modelled fields are given by 
\begin{eqnarray}
B_{n,\alpha}^m(\vec{r}_i, t; \sigma) = \sum_{j = 0}^{N_t}\left[ I_{n,\alpha}^{m(c)}(\vec{r}_i, j; \sigma) q_n^m(t - j\Delta t) + I_{n,\alpha}^{m(s)}(\vec{r}_i, j; \sigma) s_n^m(t - j\Delta t) \right].
\label{eq:bwindow_mod}
\end{eqnarray}

For ground observations (see Section \ref{sec:local}), we used
\begin{eqnarray}
I_{n,\alpha}^{m(c)}(\vec{r}_i, j; \sigma) = \int_{t_j -\Delta t/2}^{t_j +\Delta t/2} B_{n,\alpha}^{m(c)}(\vec{r}_i, \tau; \sigma)\mathrm{d}\tau, \\
I_{n,\alpha}^{m(s)}(\vec{r}_i, j; \sigma) = \int_{t_j -\Delta t/2}^{t_j +\Delta t/2} B_{n,\alpha}^{m(s)}(\vec{r}_i, \tau; \sigma)\mathrm{d}\tau.
\end{eqnarray}
Note that the two equations above are valid for magnetic fields computed either in 1-D or 3-D conductivity models.

For satellite measurements (see Section \ref{sec:global}) and a 1-D subsurface conductivity distribution, we take
\begin{eqnarray}
I_{n,\theta}^{m(c)}(\vec{r}_i,j; \sigma) &=& -I_{Q_n}(j; \sigma) \left(\frac{a}{r_i}\right)^{n+2} \frac{\mathrm{d}P_n^m(\cos\theta)}{\mathrm{d}\theta}\Big |_{\theta =\theta_i}\cos(m\phi_i), \\
I_{n,\phi}^{m(c)}(\vec{r}_i,j; \sigma) &=& I_{Q_n}(j; \sigma)\left(\frac{a}{r_i}\right)^{n+2} \frac{m}{\sin\theta_i}P_n^m(\cos\theta_i)\sin(m\phi_i), \\
I_{n,\theta}^{m(s)}(\vec{r}_i,j; \sigma) &=& -I_{Q_n}(j; \sigma) \left(\frac{a}{r_i}\right)^{n+2} \frac{\mathrm{d}P_n^m(\cos\theta)}{\mathrm{d}\theta}\Big |_{\theta =\theta_i}\sin(m\phi_i), \\
I_{n,\phi}^{m(s)}(\vec{r}_i,j; \sigma) &=& -I_{Q_n}(j; \sigma)\left(\frac{a}{r_i}\right)^{n+2} \frac{m}{\sin\theta_i}P_n^m(\cos\theta_i)\cos(m\phi_i).
\end{eqnarray}
Similar, although more lengthy, expressions can be derived using eqs. (\ref{eq:vint3d_cos})-(\ref{eq:vint3d_sin}) for a 3-D subsurface conductivity distribution.

Note that since we have eliminated internal coefficients, it suffices to use only horizontal magnetic field components in eq. (\ref{eq:minprob}) to determine inducing coefficients. This allows for more accurate description of the inducing source since horizontal components are less sensitive to the currents induced in the subsurface compared to the vertical component
\cite{Kuvshinov2008}. Since the problem is linear with respect to inducing coefficients, we used a Huber-weighted robust regression method to find the minimizer of (\ref{eq:minprob}). 

For every time window, the performance of the model can be evaluated by means of a $R^2$ statistics, called coefficient of determination. To define it, let us assume that for a given field component all observations and modeled fields in a time window $j$ are collected into vectors $\mathbf{b}_j^{\mathrm{obs}}$ and $\mathbf{b}_j^{\mathrm{mod}}$ such that
\begin{eqnarray}
\mathbf{r}_j = \mathbf{b}_j^{\mathrm{obs}} - \mathbf{b}_j^{\mathrm{mod}}
\end{eqnarray}
is the vector of residuals. Then
\begin{eqnarray}
R^2_j = 1 - \frac{\langle \mathbf{r}_j, \mathbf{r}_j \rangle}{\langle \mathbf{b}_j^{\mathrm{obs}} - \overline{b}_j^{\mathrm {obs}}, \mathbf{b}_j^{\mathrm{obs}} - \overline{b}_j^{\mathrm{\mathrm{obs}}} \rangle}
\label{eq:r2stat}
\end{eqnarray}
is the coefficient of determination for time window $j$. Here $\overline{b}_j^{\mathrm{obs}}$ denotes the mean value of $\mathbf{b}_j^{\mathrm{obs}}$ and $\langle\cdot,\cdot\rangle$ is an inner product. Note that we assumed a uniform measurement error of 1 nT when calculating $R^2$.

\subsection{Determination of induced coefficients}

Previous sections concentrated on evaluation of inducing coefficients. Once they are estimated, we can evaluate induced coefficients that describe EM fields induced in the planetary interior. This is useful in induction studies, where pairs of inducing and induced coefficients are used to estimate subsurface transfer functions, which can be ultimately inverted for the electrical conductivity distribution in the subsurface \cite{Puethe2014_Q}. 

By adopting our approach, induced coefficients can be estimated from the radial component alone. This is advantageous since the radial field exhibits higher sensitivity to the subsurface induction effects and was excluded from the estimation of the inducing coefficients (see eq. \ref{eq:minprob}). 

Provided that the inducing coefficients $q_n^m, s_n^m$ were estimated following the approach presented in the previous section, the induced part of the total magnetic field can be isolated. In particular, for the observed radial magnetic field,
\begin{equation}
    B_r^{\mathrm{int,o}}(\vec{r}, t) = B_r^{\mathrm o}(\vec{r}, t) - B_r^{\mathrm{ext,o}}(\vec{r}, t),
\end{equation}
where
\begin{equation}
    B_r^{\mathrm{ext,o}}(\vec{r}, t) =  -\sum_{(n, m) \in \mathcal{M}^+} \left[q_n^m(t)\cos (m\phi) + s_n^m(t)\sin(m\phi)  \right]n\left(\frac{r}{a}\right)^{n-1}P_n^m(\cos \theta),
\end{equation}
is the inducing part of the radial field (see eq. \ref{eq:potfield}). 

Following eq. (\ref{eq:magpotential_int}), the remaining induced part of the radial field above the ground can be expanded as
\begin{equation}
    B_r^{\mathrm{int}}(\vec{r}, t) = \sum_{k,l} \left[ g_k^l(t)\cos(k\phi) + h_k^l(t)\sin(k\phi) \right](k + 1)\left( \frac{a}{r} \right)^{k+2} P_k^l(\cos{\theta}),
\end{equation}
which is suitable for the estimation of the induced coefficients in a statistical manner. Specifically, we can estimate coefficients for a time bin centered at $t = j\Delta t$ by solving a minimization problem
\begin{equation}
\label{eq:minprob_int}
\mathbf{g}^{\ast}, \mathbf{h}^{\ast} = \argmin_{\mathbf{g}, \mathbf{h}} \sum_{i \in \mathcal{D}_t} \left[ B_{r,i}^{\mathrm{int,o}} - \sum_{k,l} \left[ g_{k,j}^l\cos(k\phi_i) + h_{k,j}^l\sin(k\phi_i) \right](k + 1)\left( \frac{a}{r_i} \right)^{k+2} P_k^l(\cos{\theta_i}) \right]^2.
\end{equation}

Therefore, by virtue of eqs. (\ref{eq:minprob}) and (\ref{eq:minprob_int}) pairs of inducing $\mathbf{q}^{\ast}, \mathbf{s}^{\ast}$ and induced $\mathbf{g}^{\ast}, \mathbf{h}^{\ast}$ coefficients can be estimated in time bins of constant length $\Delta t$, providing input data for mantle conductivity studies. Note that estimation of inducing and induced coefficients can be performed repeatedly with updated mantle conductivity models.

\section{Data}

\subsection{Geomagnetic observatories}
\label{sec:obsdata}

\begin{figure}[htbp]
  \centering
  \includegraphics[width=1.0\textwidth]{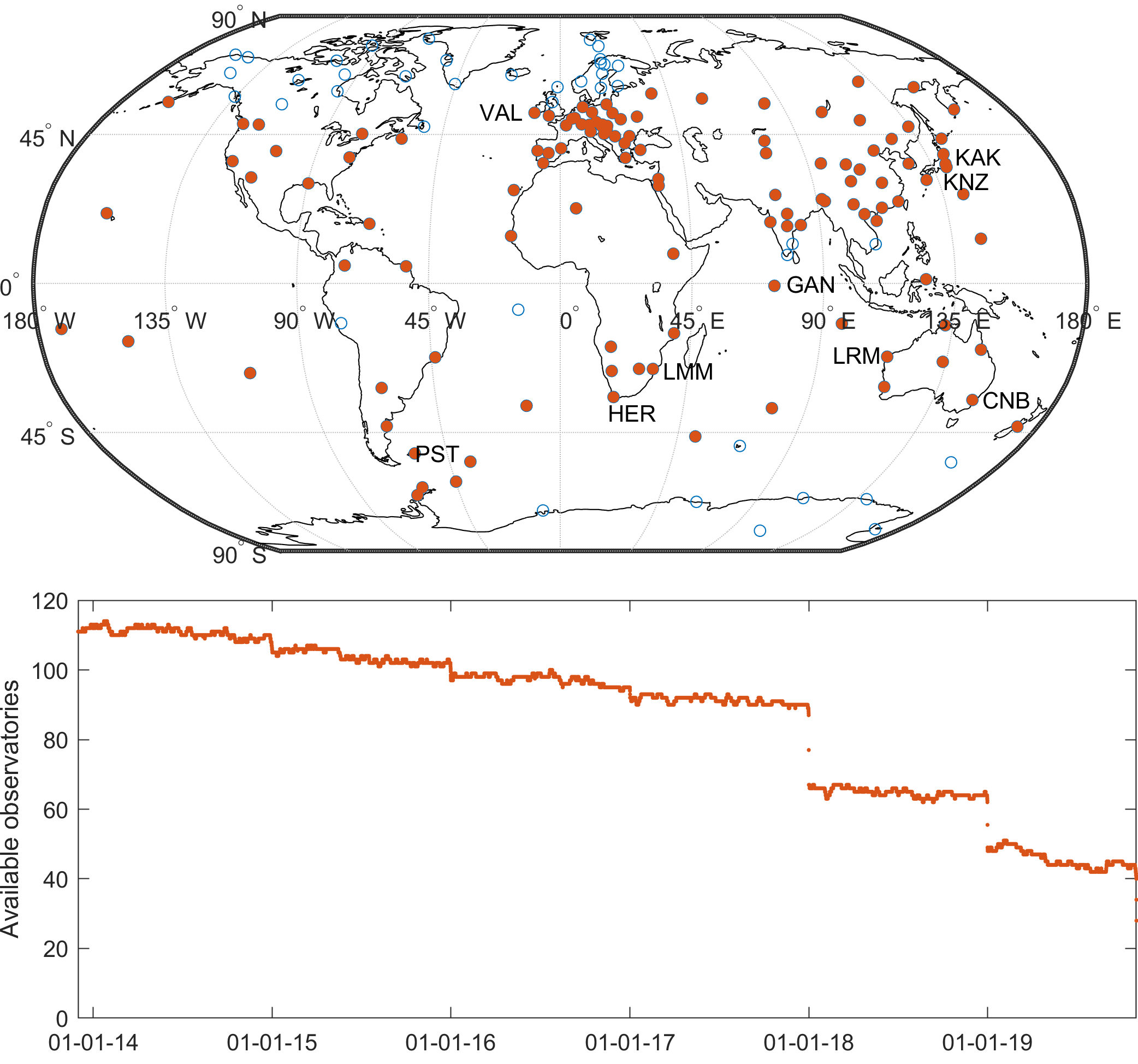}
  \caption{Top: Distribution of geomagnetic observatories. Location of geomagnetic observatories are denoted with circles. Filled circles show observatories used in this study, after discarding locations at high and equatorial geomagnetic dipole latitudes. Bottom: number of used observatories over the time period of the study.}
  \label{fig:observatories}
\end{figure}

We  apply the developed methods to the ground geomagnetic observatory data. Specifically, we took a set of quality-controlled measurements of the hourly mean vector magnetic field compiled by the British Geological Survey \cite{macmillan2013}. We concentrate here on the \textit{Swarm} era measurements by using data collected between 2013-12-01 and 2019-11-01. The model of the core and crustal fields as given by the Comprehensive Inversion (CI) model \cite{sabaka2018comprehensive} was subtracted. The distribution of the observatories over the time range used in this study is shown in Figure \ref{fig:observatories}. We further excluded observatories poleward of the $56^{\circ}$ and equatorward of $5^{\circ}$ geomagnetic latitudes. Thus, the variations in the remaining data set are predominantly driven by the mid latitude ionospheric and magnetospheric currents. The polar and equatorial latitudes are excluded because the present distribution of geomagnetic observatories can not adequately resolve spatiotemporal structures of the dominant current systems at these latitudes. 

\subsection{Geomagnetic satellites}
\label{sec:satdata}

We used nearly six years (2013-12-01 -- 2019-11-01) of the geomagnetic field measurements taken by the \textit{Swarm} Alpha and Bravo satellites. Similar to the observatory data, core and crustal fields as given by the CI model were subtracted. The time windows of three hours were used, which corresponds to two full orbits and aims to improve the data coverage within a window. Here, we concentrate on studying the EM induction effects of the large-scale magnetosphere currents of external origin and thus the day side data, namely between 5 AM and 7 PM local time, was excluded. 

\section{Results}

\subsection{Transfer functions and impulse responses}

All transfer functions and corresponding impulse responses referred to as "1-D" were calculated by taking a conductivity model that consists of the 1-D conductivity profile from \citeA{grayver2017} with a 7000 S conductance layer that represents average conductance of the oceans and sediments. For the results referred to as "3-D", a laterally heterogeneous conductivity shell of $1/4^{\circ}$ resolution was used to account for the variations in the ocean bathymetry and thickness of sediments. For the 1-D case, transfer function were calculated analytically, whereas 3-D transfer functions were calculated numerically by solving Maxwell's equations in a spherical shell with a Finite Element code GoFEM \cite{grayver2015large, grayver2019, dealII92}.

\begin{figure}
  \centering
  \includegraphics[width=1.0\textwidth]{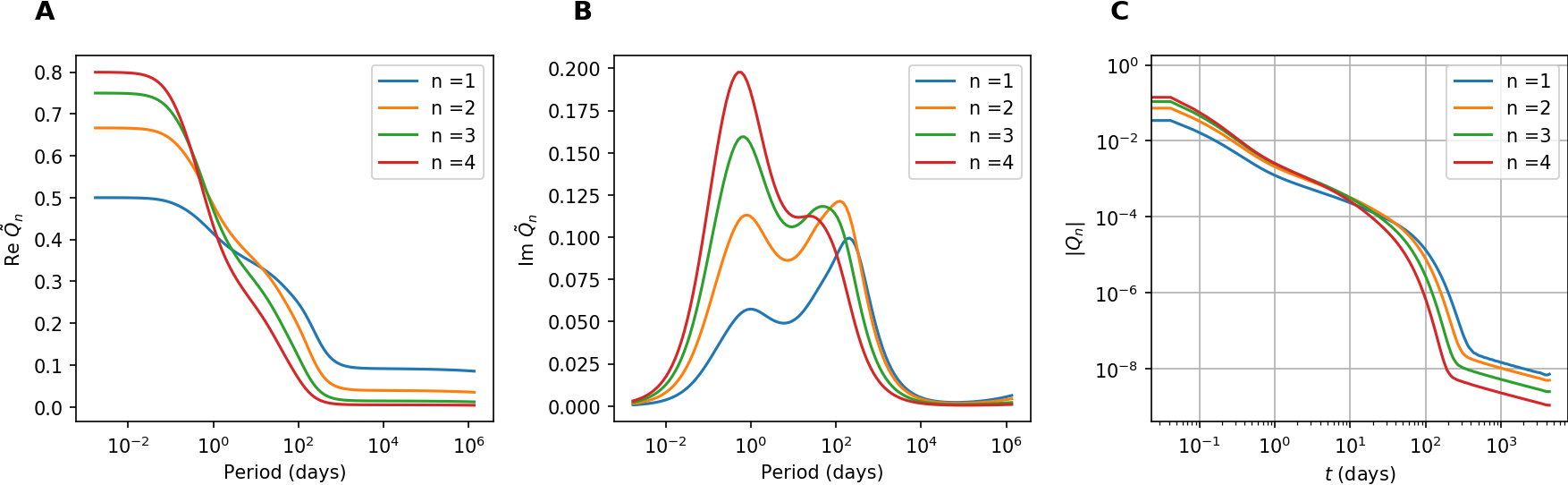} 
  \caption{Real (A) and imaginary (B) parts of the $\tilde{Q}_n$ transfer functions (eq. \ref{eq:qresponse}) for different degrees $n$ and 1-D conductivity profile of \citeA{grayver2017}. The magnitudes of the corresponding discrete impulse responses (eq. \ref{eq:qn_discrete}) are shown in plot (C).}
  \label{fig:qkernels1D}
\end{figure}

Figure \ref{fig:qkernels1D} shows 1-D transfer functions and corresponding discrete impulse responses. As expected, we see that the decay rate for responses with higher degrees $n$ is faster, implying that attenuation rate of the induced currents increases with the SH degrees of the inducing field. At periods of $1$ year and longer, real part of the transfer function flattens as a result of the transient induction effect of the core, which has a finite conductivity \cite{velimsky2003transient}.

\begin{figure}
  \centering
  \includegraphics[width=1.0\textwidth]{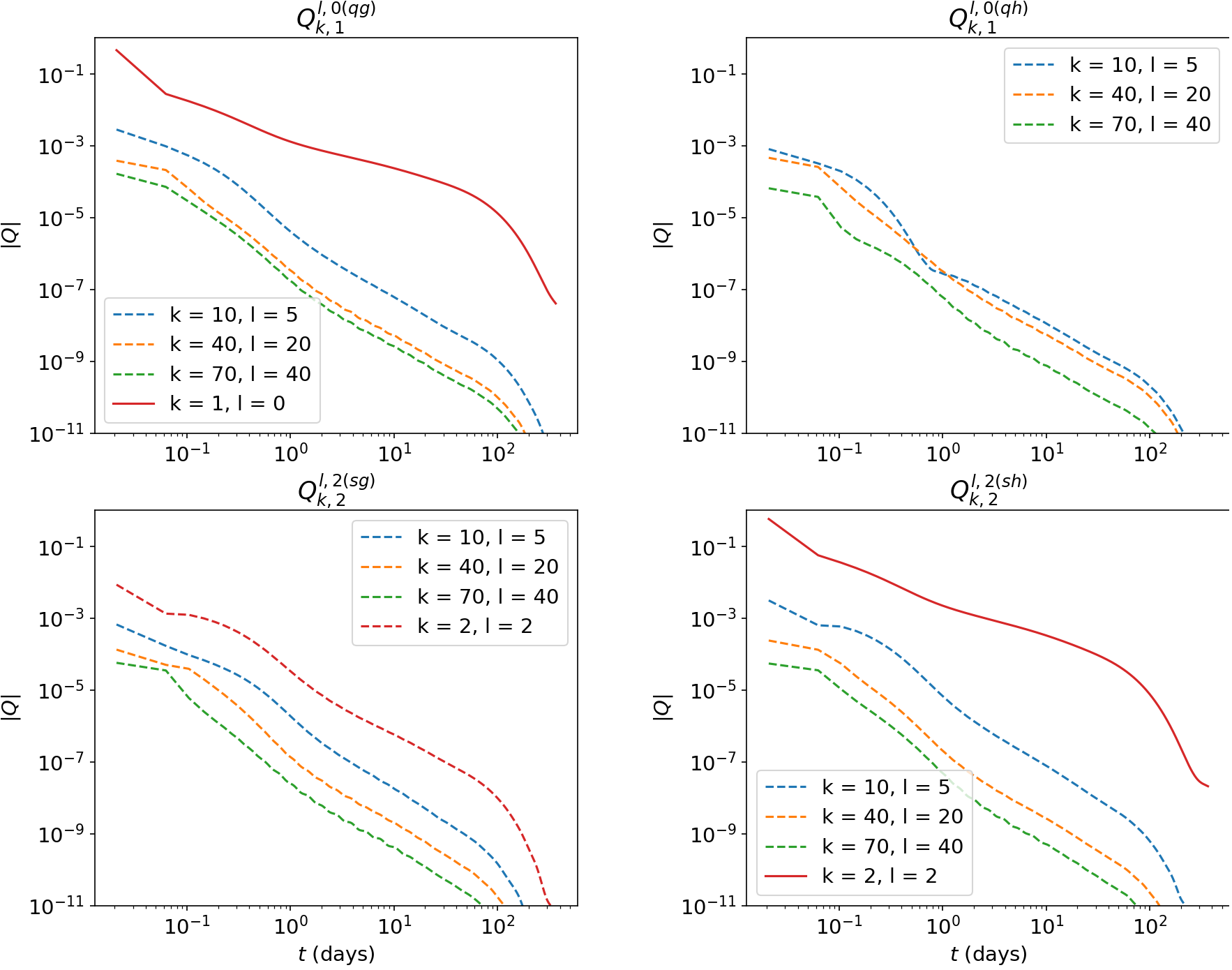} 
  \caption{A selection of the 3-D discrete impulse responses from the $Q_{kn}^{lm,qg}$ and $Q_{kn}^{lm,qh}$ matrices (eqs. \ref{eq:vint3d_cos}-\ref{eq:vint3d_sin}) due to the $q_1^0$ (top row) and $s_2^2$ (bottom row) inducing terms. Dashed lines denote responses which are non zero only in the case of a 3-D conductivity distribution.}
  \label{fig:qkernels3D}
\end{figure}

Figure \ref{fig:qkernels3D} shows a set of discrete impulse responses from the 3-D $Q$-matrix for different external and internal degrees and orders. First of all, note that in 3-D, the matrix is dense, i.e. each inducing coefficients leads to infinitely many induced coefficients. However, we observe that the diagonal elements dominate the matrix, whereas off-diagonal entries are generally smaller and decay with the SH degree. 

\begin{figure}
  \centering
  \includegraphics[width=1.0\textwidth]{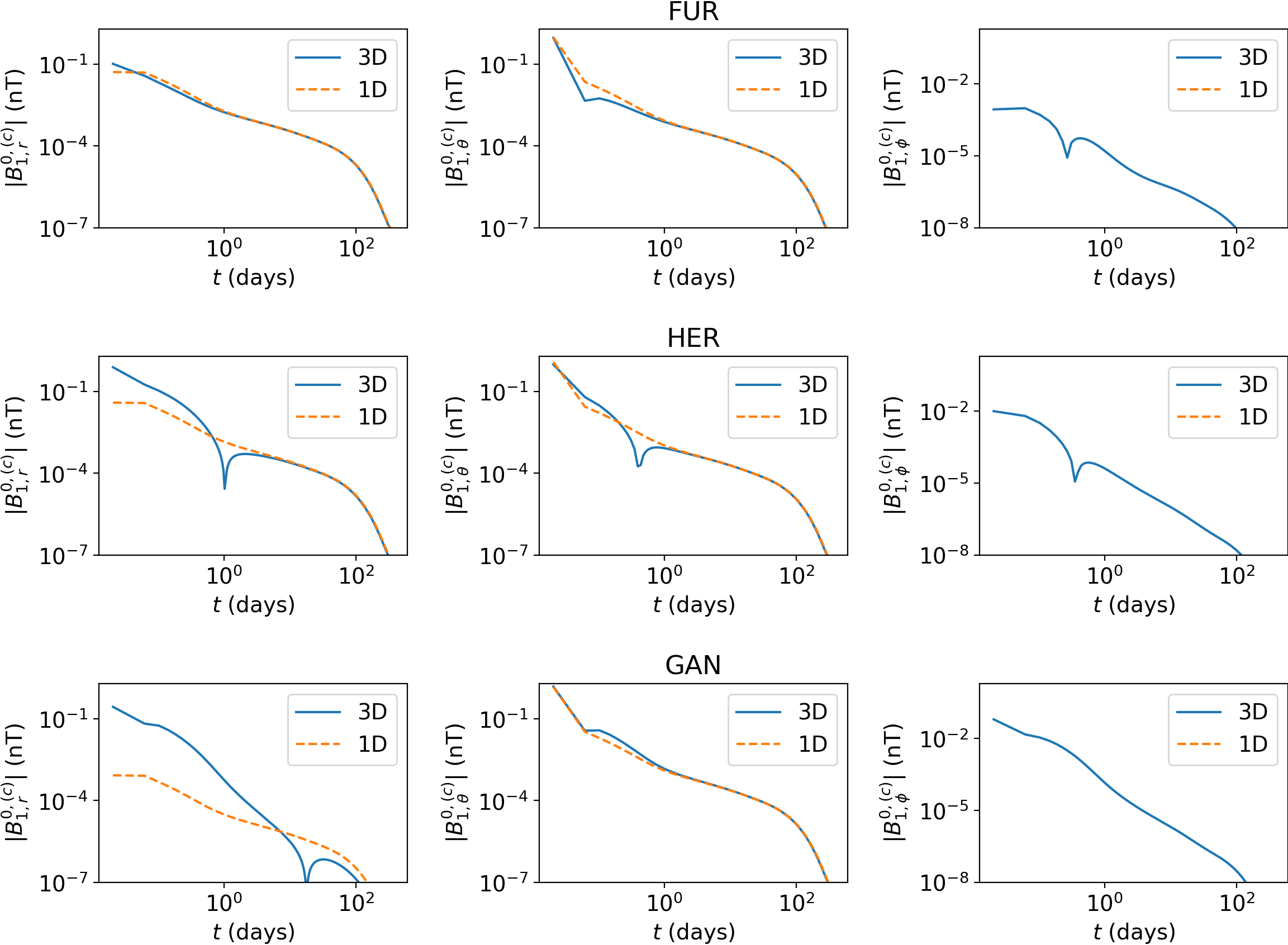} 
  \caption{Magnetic field discrete impulse responses (eq. \ref{eq:bsin_cos}) due to a $q_1^0$ inducing field for three magnetic field components (columns) at three locations: F\"urstenfeldbruck (FUR), Hermanus (HER) and Gan, Maldives (GAN). Both 1-D (dashed lines) and 3-D responses (solid lines) are shown.}
  \label{fig:bkernels}
\end{figure}

Finally, Figure \ref{fig:bkernels} shows examples for local impulse responses at several observatory locations where both 1-D and 3-D responses are plotted to highlight the effect of the ocean and sedimentary cover on impulse responses. We see that the difference between 1-D and 3-D responses is particularly large for island and coastal locations. 





\subsection{Model of external magnetic field variations from ground observations}
\label{sec:obs_model}

We determined SH coefficients up to degree $n_{\mathrm{max}} = 3$ and order $m_{\mathrm{max}} = 3$ within hourly time bins. The length of impulse responses was set to six months, thus transient effects older than six months are neglected. This choice is justified since impulse responses for time lags larger than six months are $\leq 10^{-6}$ (see Figures \ref{fig:qkernels1D}-\ref{fig:bkernels}), thus the transient effects become negligible for the  majority of practical applications. Other details pertained to data pre-processing and the method of evaluating SH coefficients are given in Sections \ref{sec:obsdata} and \ref{sec:ext_coeffs}, respectively.

\begin{figure}
  \centering
  \includegraphics[width=0.8\textwidth]{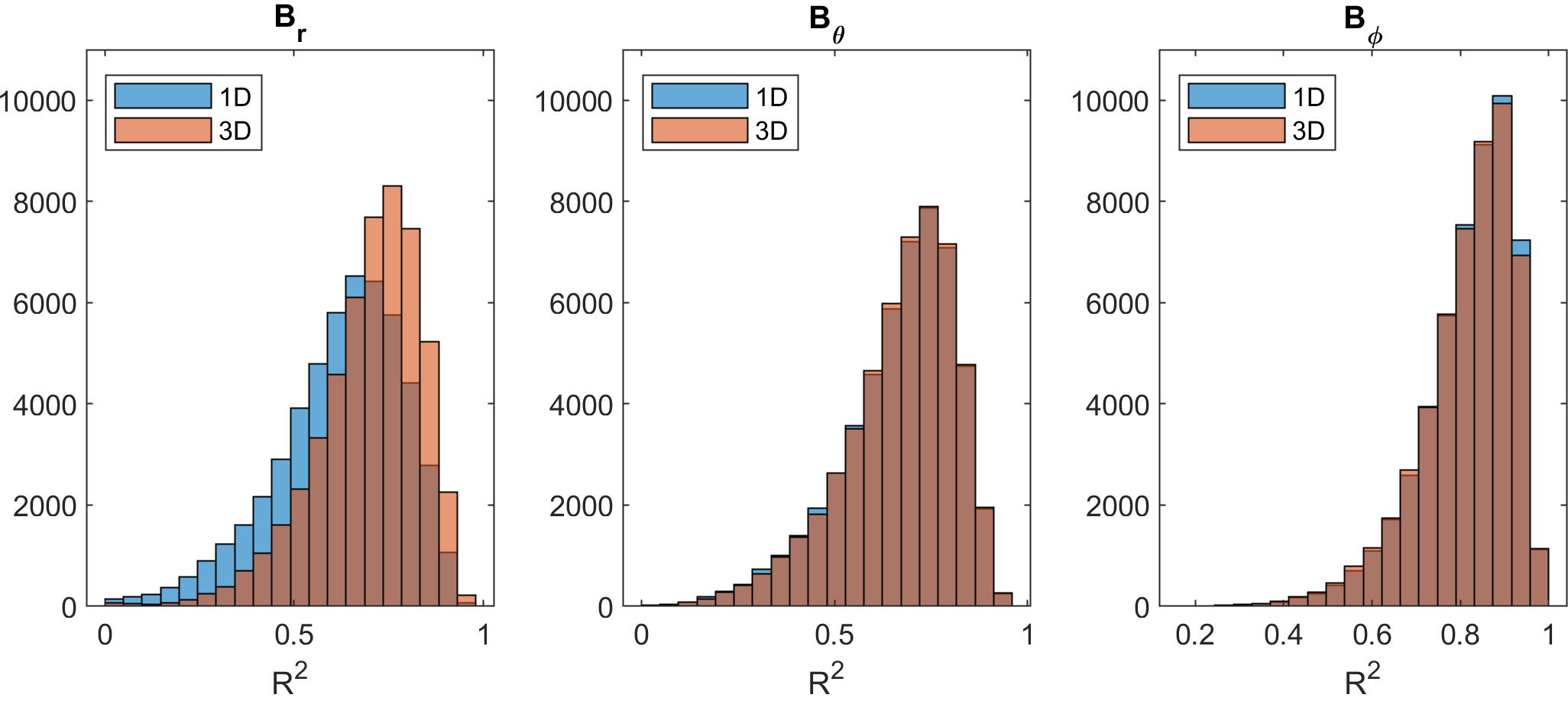} 
  \caption{Histograms of the $R^2$ statistics (coefficient of determination) for individual magnetic field components (from left to right: $B_r, B_{\theta}, B_{\phi}$) and all time windows. The $R^2$ statistics was determined following eq. (\ref{eq:r2stat}) between observatory data and model predictions. The model details are described in Section \ref{sec:obs_model}.}
  \label{fig:r2_obs}
\end{figure}

The coefficients were determined using both 1-D and 3-D impulse responses from horizontal magnetic field components ($B_{\theta}, B_{\phi}$). Subsequently, coefficient of determination $R^2$ was calculated for every time bin using eq. (\ref{eq:r2stat}) and three components separately, including $B_r$ component, which was not used for the model construction. Figure \ref{fig:r2_obs} shows histograms of $R^2$ coefficient for 1-D and 3-D models. One apparent observation is the significantly better fit of the radial component with a 3-D conductivity model. The fit for horizontal components is virtually identical, the differences are minute and likely fall within the modelling and observation errors. Noteworthy that among all components, the highest coherency is observed for the longitudinal component. These observations confirm that our model, especially the one based on a 3-D conductivity model, has a predictive power.

\begin{figure}
  \centering
  \includegraphics[width=0.8\textwidth]{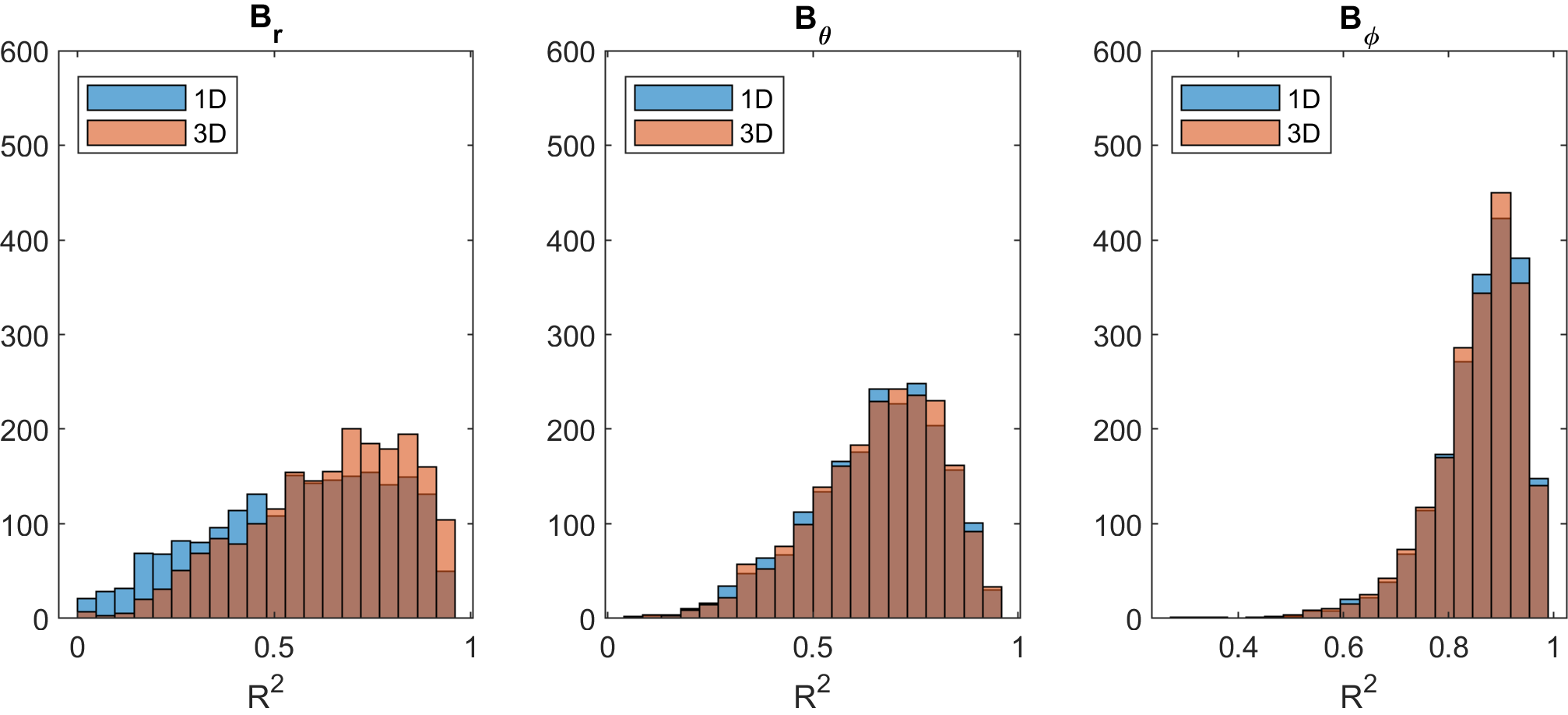} 
  \caption{Same as Figure \ref{fig:r2_obs}, but restricted to time windows when $|Dst| > 40$ nT.}
  \label{fig:r2_obs_dst}
\end{figure}

To test how the model performs during different magnetic conditions, we further plot histograms of $R^2$ statistics for times when magnetic variations are dominated by magnetospheric disturbances (here defined as $|Dst| > 40$ nT) in Figure \ref{fig:r2_obs_dst}. Although we still observe a significant improvement in coherency for $B_r$, generally the correlation is lower for $B_r, B_{\theta}$, whereas it remains high for the longitudinal component. Further, we plot $R^2$ histograms for times when $Kp \leq 2$. The reason to use $Kp$ instead of $Dst$ this time is to emphasize the quiet ionosphere conditions. Similar to the examples with disturbed magnetosphere, we again observe significant improvements in the radial component for the 3-D model. In comparison with the previous case, however, we see systematically higher $R^2$ values for all components. Therefore, our model exhibits a better fit during quiet times. Further, the improved fit of the $B_r$ enabled by the 3-D model is observed for all times and magnetic conditions, indicating that proper inclusion of the ocean effect is essential when modeling both magnetospheric and ionospheric variations.

\begin{figure}
  \centering
  \includegraphics[width=0.8\textwidth]{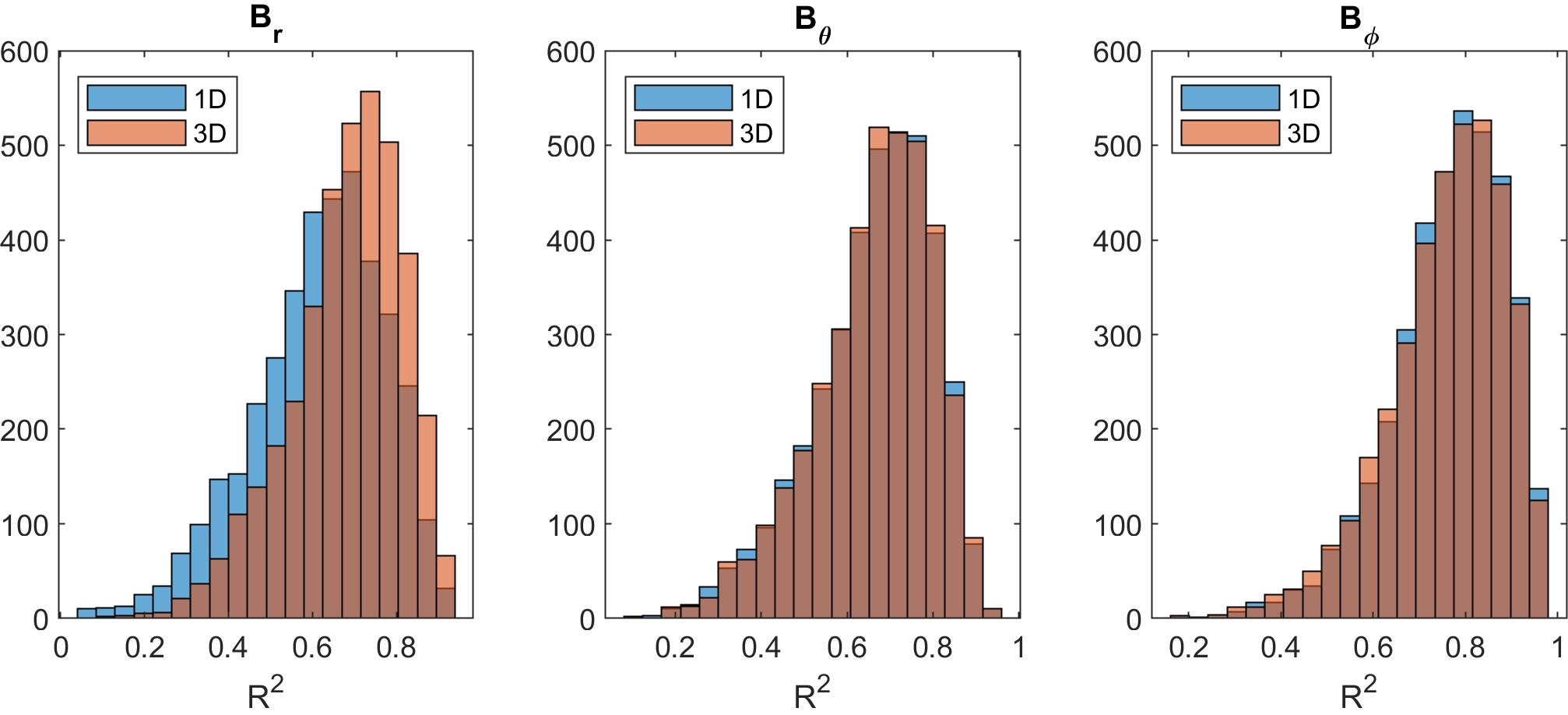} 
  \caption{Same as Figure \ref{fig:r2_obs}, but restricted to time windows when $Kp \leq 2$.}
  \label{fig:r2_obs_kp}
\end{figure}

\begin{figure}
  \centering
  \includegraphics[width=.9\textwidth]{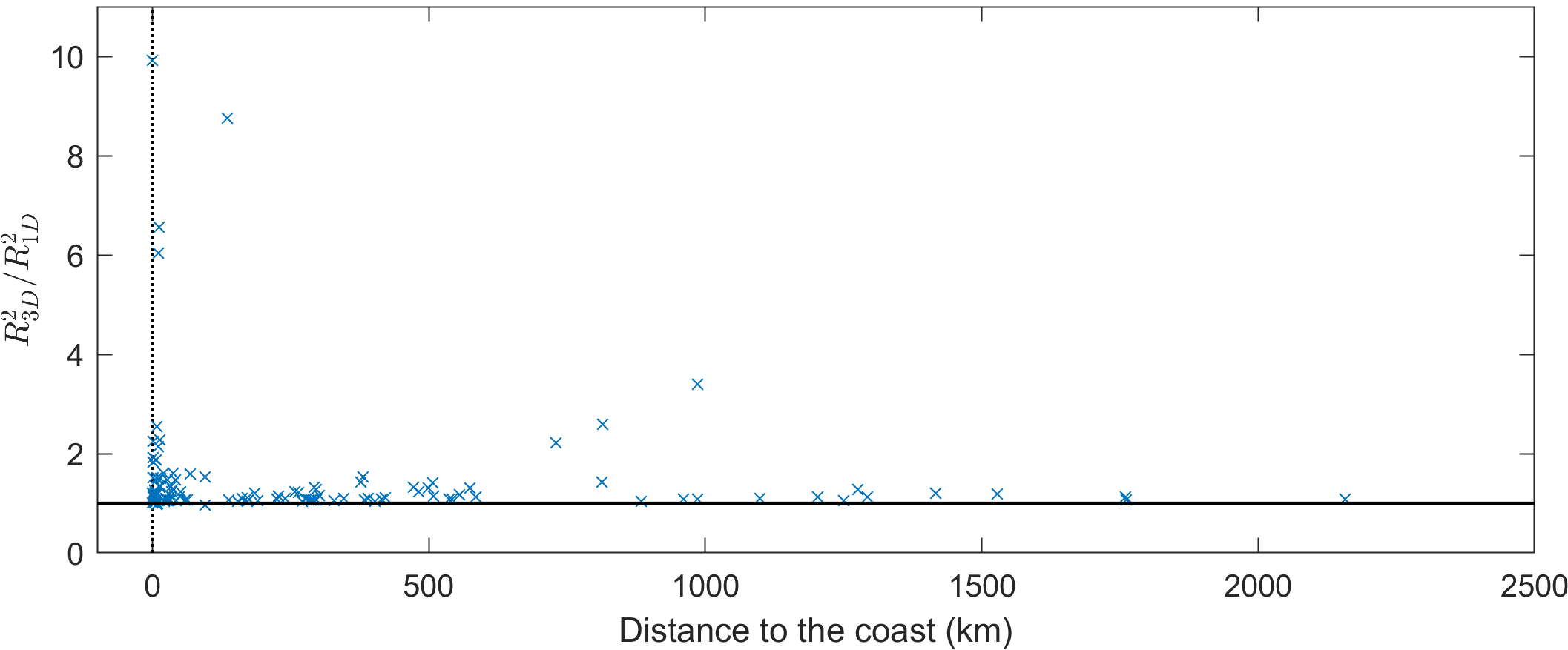} 
  \caption{Ratio of the 3-D to 1-D models $R^2$ coefficients for $B_r$ field at individual observatories plotted versus distance to the shoreline. Values larger than one indicate improvement over the 1-D model.}
  \label{fig:r2_vs_distcoast}
\end{figure}

To better quantify the effect of the improved fit due to the usage of a 3-D model, we calculated the ratio of 3-D and 1-D $R^2$ values for radial magnetic field component at all observatory locations. These values, plotted as a function of the distance to the shoreline, are shown in Figure \ref{fig:r2_vs_distcoast}. We observe the improved fit at virtually all locations with the most significant improvement up to a factor of 11 for observations that are $\leq 200$ km from the coast. However, even locations as far as 3000 km exhibit considerably better fit. This is explained by including the conductance of continental sediments in our 3-D model.

\begin{figure}
  \centering
  \includegraphics[width=\textwidth]{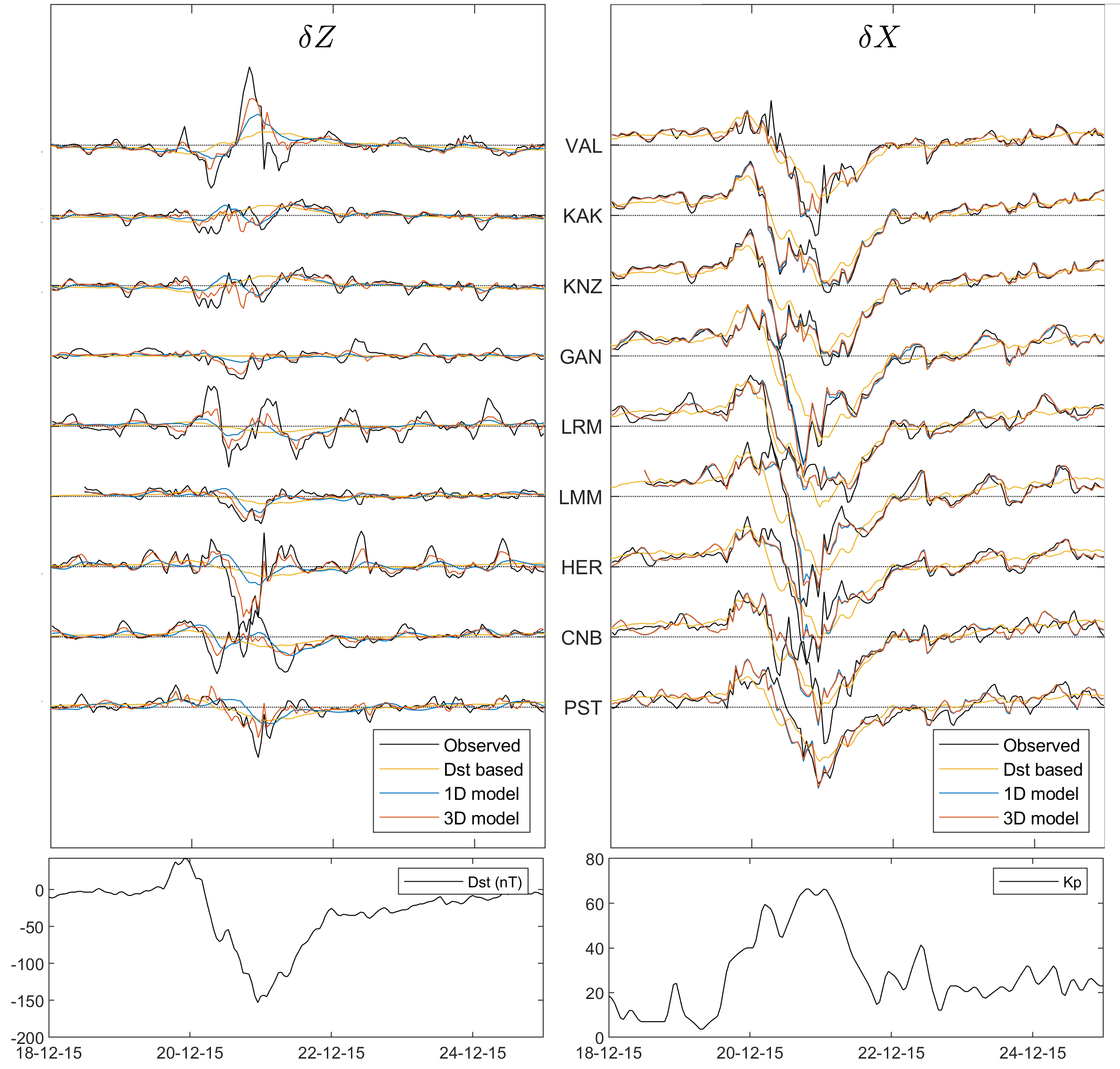}
  \caption{Time series of observed and modelled variations in horizontal ($\delta X = -\delta B_{\theta}$) and radial ($\delta Z = -\delta B_r$) components at a set of observatories, ordered by latitude. Predictions based on 1-D and 3-D conductivity models are shown along with $D_{st}$-based fields (eq. \ref{eq:Dst}). The offset between dotted lines is 100 nT. Lower panels show corresponding $D_{st}$ and Kp indices. See Figure \ref{fig:observatories} for locations of selected observatories.}
  \label{fig:Bmod_17Dec2015}
\end{figure}

\begin{figure}
  \centering
  \includegraphics[width=\textwidth]{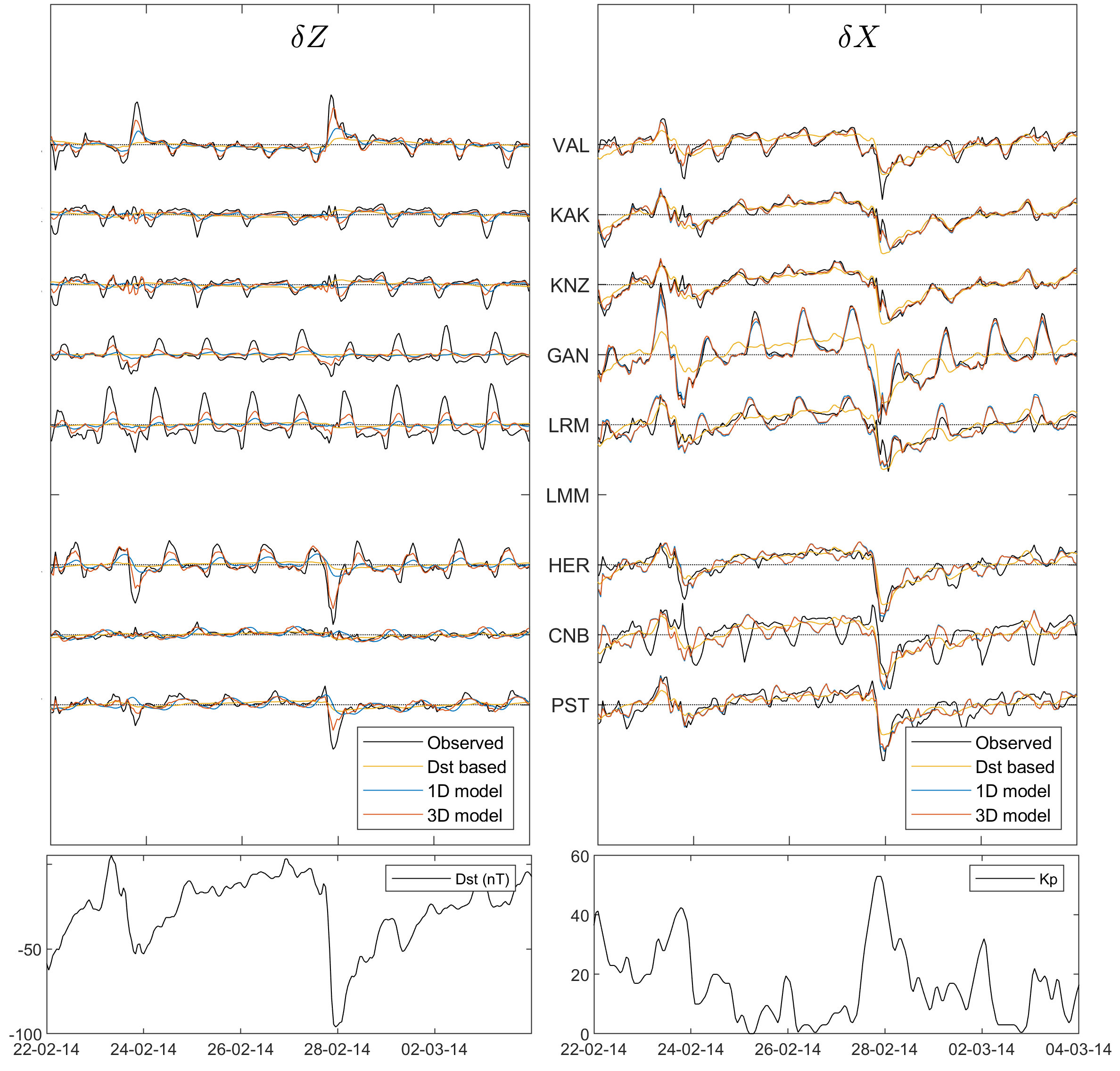}
  \caption{Same as Figure \ref{fig:Bmod_17Dec2015}, but for a different time period.}
  \label{fig:Bmod_21Feb2014}
\end{figure}

Finally, we inspect the observed and modeled time series at a selection of coastal and island observatories. Here, we also added predictions based on the Dst index, calculated as
\begin{eqnarray}
\label{eq:Dst}
B_r^{\mathrm{Dst}}(\vec r, t) &=& \big [\mathrm{Est}(t) - 2\mathrm{Ist}(t) \big ]\cos \theta \\
B_{\theta}^{\mathrm{Dst}}(\vec r, t) &=& -\big [\mathrm{Est}(t) + \mathrm{Ist}(t)\big ]\sin\theta \\
B_{\phi}^{\mathrm{Dst}}(\vec r, t) &=& 0,
\end{eqnarray}
where $\mathrm{Dst}(t) = \mathrm{Est}(t) + \mathrm{Ist}(t)$ is a sum of inducing and induced terms \cite{maus2004, olsen2005}. Figures \ref{fig:Bmod_17Dec2015}-\ref{fig:Bmod_21Feb2014} each show one week of observed magnetic field variations and model predictions. These periods were chosen since they cover both magnetically disturbed and quiet conditions. 

The origin of the discrepancy in amplitude between the observed and modelled fields is twofold: (i) we used the global average mantle conductivity profile whereas in reality the bulk subsurface conductivity varies laterally; and (ii) slightly larger discrepancy for quiet times indicates that the adopted SH parameterization with $n_{\mathrm{max}} = 3, m_{\mathrm{max}} = 3$ is still insufficient to explain these variations, mostly related to ionospheric currents \cite{guzavina2019probing, schmucker1999spherical}.

\begin{figure}
  \centering
  \includegraphics[width=\textwidth]{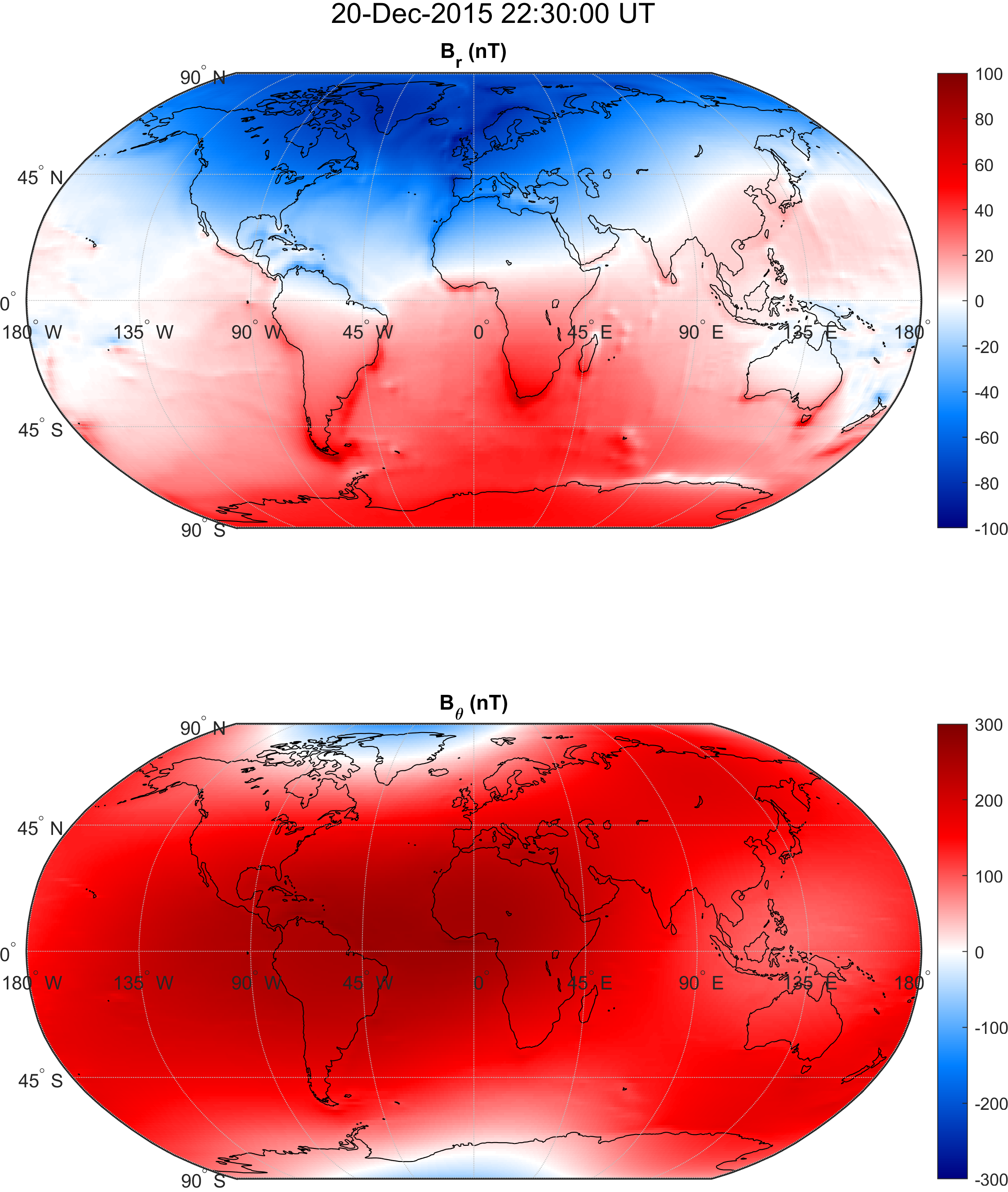}
  \caption{Maps of the radial (top) and horizontal (bottom) components of modelled magnetic field variations at a surface. Predictions based on the 3-D conductivity model for a given UT instance are shown. The recorded $D_{st}$ index value at this instance was -155 nT. Significant coastal EM induction effects are visible in the radial component.}
  \label{fig:B_3В_mod_20151220}
\end{figure}

\begin{figure}
  \centering
  \includegraphics[width=\textwidth]{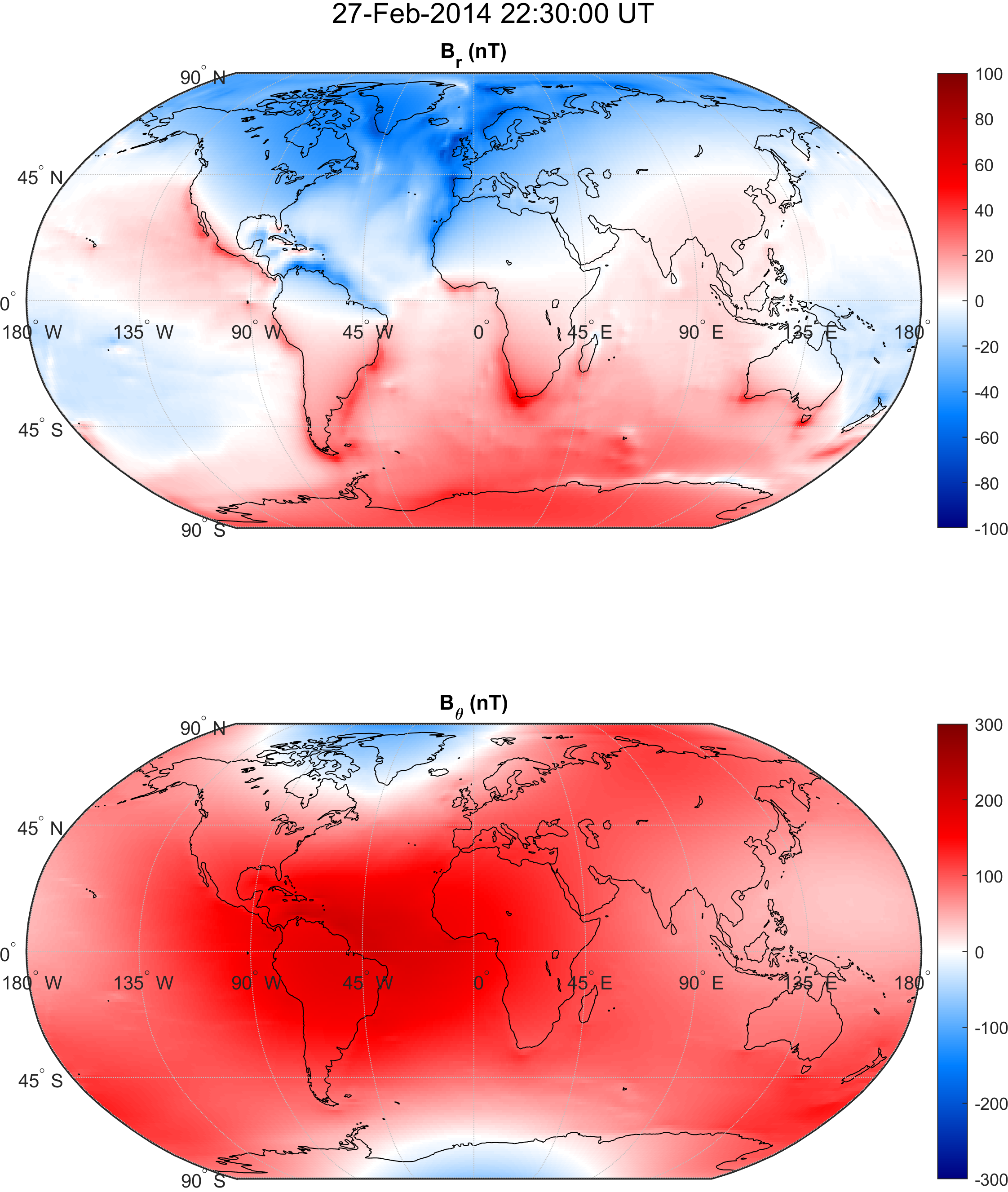}
  \caption{Same as Figure \protect \ref{fig:B_3В_mod_20151220}, but for a different time. The $D_{st}$ index value at this instance was -95 nT.}
  \label{fig:B_3В_mod_20140227}
\end{figure}

Additionally, Figures \protect \ref{fig:B_3В_mod_20151220}-\ref{fig:B_3В_mod_20140227} show spatial distribution of the magnetic field as predicted by estimated external coefficients and a 3-D conductivity model. Much stronger influence of 3-D EM induction effects in the $B_r$ components are clearly visible. Most of these effects occur near coastal areas and strong lateral conductivity gradients.

\subsection{Model of magnetospheric ring current variations from \textit{Swarm} observations}
\label{sec:sat_model}

In this section, the model of inducing coefficients was determined by using only satellite data, which was described in Section \ref{sec:satdata}. Since we work with night-side data and two satellites, we determined SH coefficients up to degree $n_{\mathrm{max}} = 2$ and order $m_{\mathrm{max}} = 1$ using time bins of 3 hours. Therefore, the resolution of this model is much lower than the model in previous section that was based on observatory data. Other parameters pertained to data pre-processing and evaluation of SH coefficients are described in Sections \ref{sec:satdata} and \ref{sec:ext_coeffs}, respectively.

\begin{figure}
  \centering
  \includegraphics[width=\textwidth]{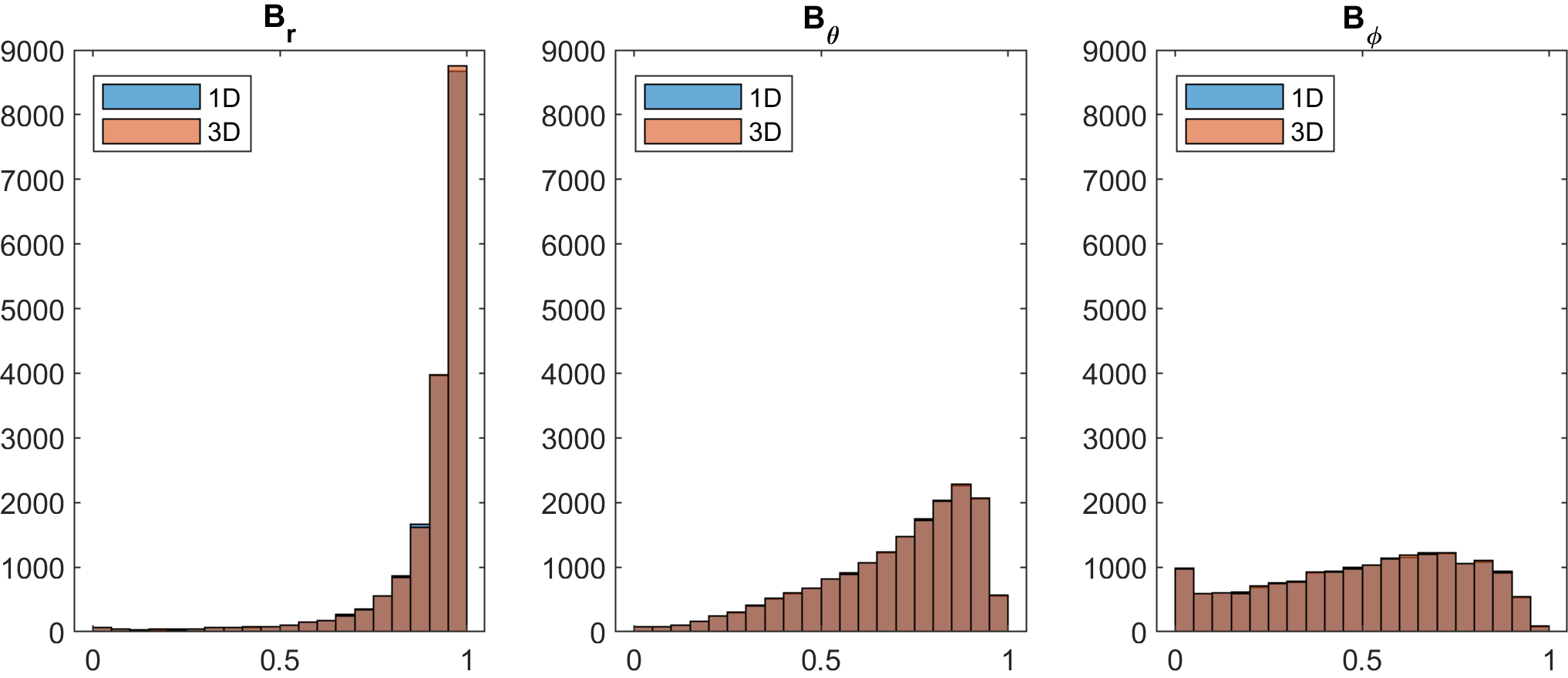}
  \caption{Histograms of the $R^2$ statistics (coefficient of determination) for individual magnetic field components and all time windows. The $R^2$ statistics was determined following eq. (\ref{eq:r2stat}) between satellite measurements and predictions based on the model described in Section \ref{sec:sat_model}.}
  \label{fig:r2_1d_vs_3d_sat}
\end{figure}

\begin{figure}
  \centering
  \includegraphics[width=\textwidth]{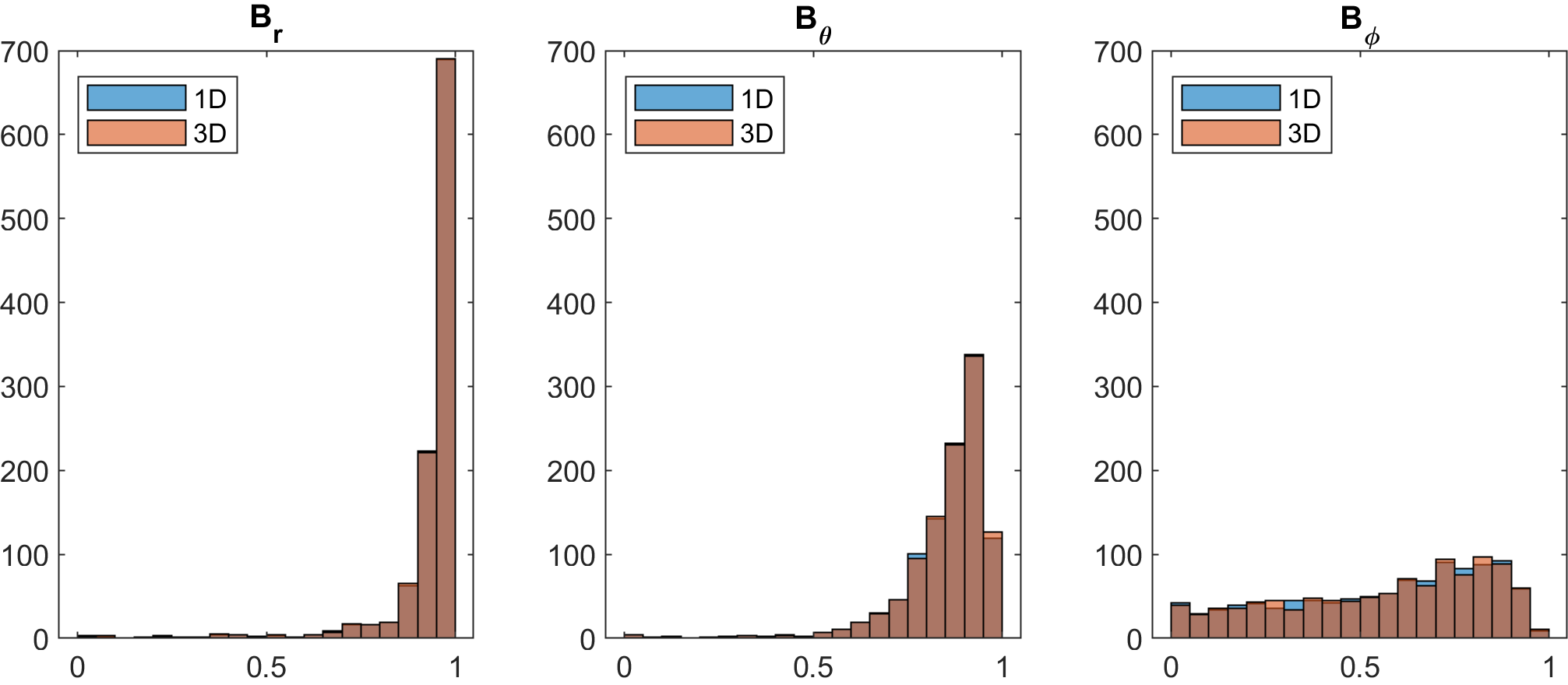}
  \caption{Same as Figure \ref{fig:r2_1d_vs_3d_sat}, but restricted to time windows when $Kp \leq 2$.}
  \label{fig:r2_1d_vs_3d_sat_quiet}
\end{figure}

As in the previous section, we first look at the distribution of $R^2$ statistics for all time bins and magnetic field components (see Figure \ref{fig:r2_1d_vs_3d_sat}). First observation that we make is that $R^2$ values are very similar between 1-D and 3-D models, indicating that 3-D induction effect from the ocean is largely attenuated at satellite altitudes. Interestingly that now we also have much higher $R^2$ values for the radial component compared to the $B_{\theta}$, even though $B_r$ was not used in the construction of the model. One possible explanation are signals that mostly affect horizontal ($B_{\theta}$, $B_{\phi}$) components at mid latitudes, such as those generated by F-region ionospheric currents \cite{olsen1997ionospheric}. These signals cannot be explained by our low-resolution parameterization that is based on the potential field assumption. To test this hypothesis, histograms limited to the time windows for which $Kp \leq 2$ are shown in Figure \ref{fig:r2_1d_vs_3d_sat_quiet}. Indeed, during periods with the less disturbed ionosphere we obtain significantly higher values of $R^2$ for the $B_{\theta}$ component.

\begin{figure}
  \centering
  \includegraphics[width=\textwidth]{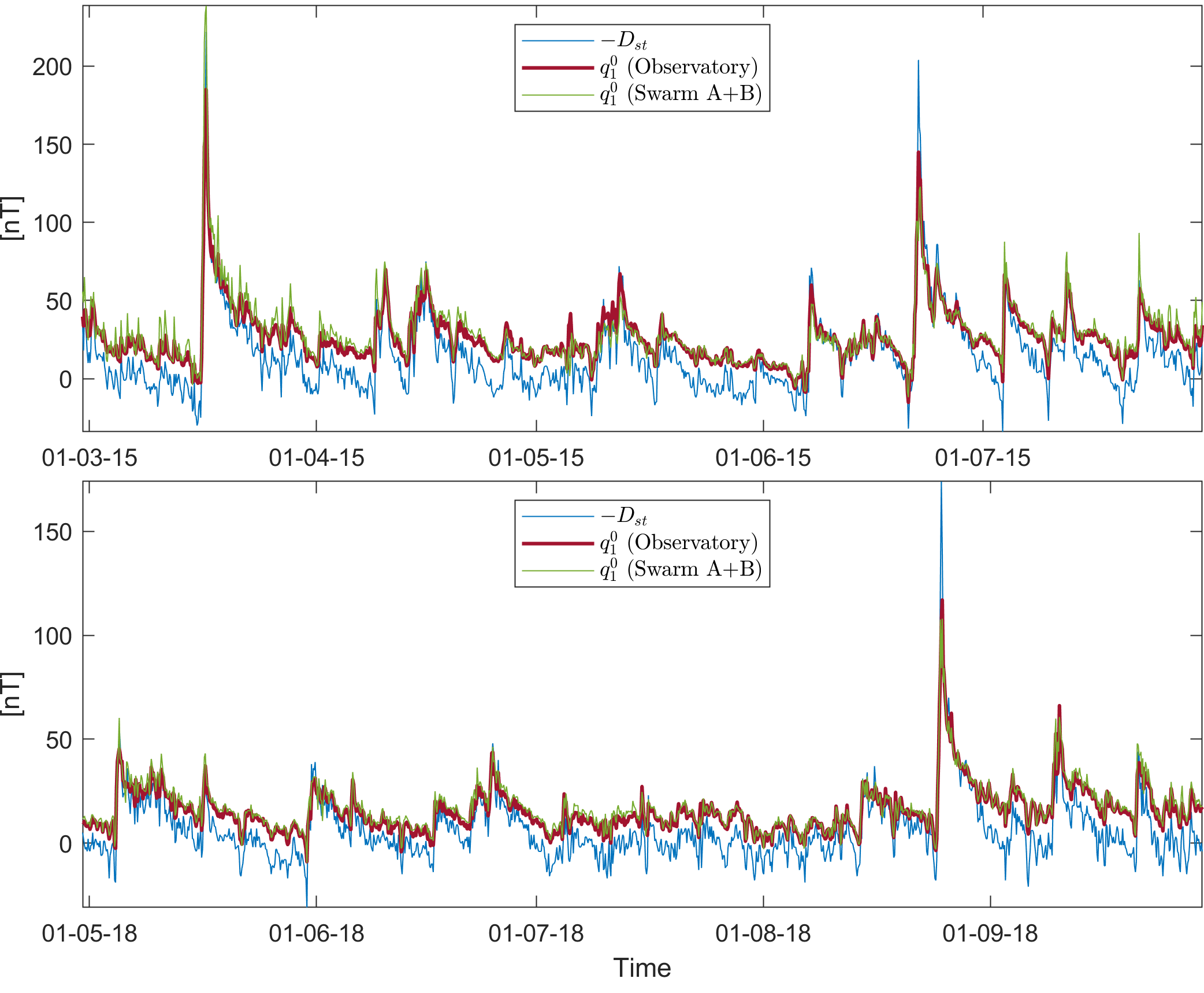}
  \caption{Time series of the first zonal SH coefficient, $q_1^0$, as given by satellite (Section 4.3) and observatory (Section 4.2) data based models in geomagnetic coordinate frame. Two five months intervals featuring quiet and disturbed magnetic conditions are shown. For comparison, the negative $D_{st}$ magnetic index is plotted. Systematic offset in $D_{st}$ against $q_1^0$ seen in Figures \ref{fig:q10_obs_vs_sat} is due to the absence of stable quiet time ring current in the $D_{st}$ index.}
  \label{fig:q10_obs_vs_sat}
\end{figure}

Finally, Figure \ref{fig:q10_obs_vs_sat} plots time series of the $q_1^0$ coefficient determined using the observatory and satellite data. For reference, we also plot the $D_{st}$ index. We observe very good match between coefficients estimated from satellite and observatory data, confirming the validity of both models and approaches. 

\section{Conclusions}

The EM induction effect from a time-varying magnetic field significantly influences magnetic field observations, where it can be both a polluting signal to be removed or a primary signal to study (e.g. mantle induction and space weather applications). We showed that the inducing currents of ionospheric and magnetospheric origin can be effectively estimated while the effect of the planetary induced response is modeled. This work has presented a unified framework for modeling EM induction effects in ground and satellite data by means of time domain impulse responses due to arbitrary external sources and in presence of a 3-D subsurface conductivity distribution. This approach is amenable to integrate with models that involve constantly augmented time series and require "on the fly" updates of geomagnetic models.

We have elaborated the underlying mathematical machinery for the case when basis functions used for spatial parameterization of magnetic field are given by spherical harmonic functions. This choice was made owing to the ubiquity of SH basis in Earth's and planetary magnetism community. However, the approach is general and straightforward to extend to other basis functions should practical applications demand this.

We further showed that the effects from heterogeneity in subsurface electrical conductivity can dominate the radial magnetic field component and should be accounted for provided that some knowledge about 3-D subsurface conductivity structure for Earth is available. Contrary to the common presumption, the 3-D effects are significant during both quiet and disturbed magnetic conditions since the induction effect is transient, hence widely used selection criteria based on instant values of magnetic indices and local time can not completely eliminate the effects of EM induction, rendering the modeling approach presented here a suitable alternative that accounts for its transient nature. 

%
%
%
%
\appendix

\section{Properties of transfer functions and impulse responses}
\label{sec:imp_resp}

Convolution integrals such as in eqs. (\ref{eq:field_via_kernels}) and (\ref{eq:qg_td})-(\ref{eq:qs_td}) represent a response of a medium to a time-varying extraneous current. These relations follow from the (often omitted) properties of a physical system that we model. We state these properties here and discuss implications. Our presentation closely follows a more detailed analysis by \citeA{svetov1991}. 

\begin{enumerate}

\item \textbf{Linearity} allows us to define a response, $\zeta(t)$, of a medium at time $t$ to an extraneous forcing as
\begin{equation}
\label{eq:response}
    \zeta(t) = \int_{-\infty}^{\infty} \mathcal{F}(t, t') \chi(t') \mathnormal{d}t',
\end{equation}
where $\chi$ is the extraneous forcing that depends on time $t'$ and $\mathcal{F}(t, t')$ is the medium Green's function that does not depend on the amplitude of the exerted force. 

\item \textbf{Stationarity} implies that the response of a medium does not depend on the time of occurrence of the excitation. In this case $\mathcal{F}(t, t') \equiv f(t - t')$ and eq. (\ref{eq:response}) can be rewritten as a convolution integral
\begin{equation}
\label{eq:response_conv}
    \zeta(t) = \int_{-\infty}^{\infty} f(t - \tau) \chi(\tau) \mathnormal{d}\tau = \int_{-\infty}^{\infty} f(\tau) \chi(t-\tau) \mathnormal{d}\tau,
\end{equation}
where $f(t)$ represents the impulse response of a medium. In frequency domain, the convolution integral reduces to
\begin{equation}
\label{eq:response_fd}
\tilde{\zeta}(\omega) = \tilde{f}(\omega)\tilde{\chi}(\omega),
\end{equation}
where $\tilde{f}(\omega)$ is called the transfer function and we used tilde sign ($\tilde{\cdot}$) to denote complex quantities. Eqs. (\ref{eq:response_conv}) and (\ref{eq:response_fd}) are related through the Fourier transform
\begin{equation}
\label{eq:fft_ir}
\tilde{f}(\omega) = \int_{-\infty}^{\infty} f(t)e^{\mathrm{i}\omega t}\textnormal{d}t.
\end{equation}

\item Since we work in time domain with a real valued forcing, the impulse response is also \textbf{real}. To see implications of this, let us define the inverse Fourier transform of $\tilde{f}(\omega) = f_R(\omega) + \mathrm{i} f_I(\omega)$ as
\begin{eqnarray}
\label{eq:ifft_ir}
f(t) &=& \frac{1}{2\pi} \int_{-\infty}^{\infty} \tilde{f}(\omega)e^{-\mathrm{i}\omega t}\textnormal{d}\omega \nonumber \\ 
&=& \frac{1}{2\pi} \int_{-\infty}^{\infty} \left[ f_R(\omega)\cos(\omega t) + f_I(\omega)\sin(\omega t) \right]\textnormal{d}\omega \nonumber \\
&+& \frac{\mathrm{i}}{2\pi} \int_{-\infty}^{\infty} \left[ f_I(\omega)\cos(\omega t) - f_R(\omega)\sin(\omega t) \right]\textnormal{d}\omega.
\end{eqnarray}
For an impulse response to be real, the last term in the integral (\ref{eq:ifft_ir}) has to vanish. This is possible only if $f_R(\omega)$ and $f_I(\omega)$ are even and odd functions of the angular frequency $\omega$, respectively. Therefore, eq. (\ref{eq:ifft_ir}) reduces to
\begin{eqnarray}
\label{eq:ifft_ir_1}
f(t) &=& \frac{1}{\pi} \int_0^{\infty}\left[ f_R(\omega)\cos(\omega t) + f_I(\omega)\sin(\omega t) \right]\textnormal{d}\omega.
\end{eqnarray}

\item Impulse response is \textbf{causal}. This property implies that $f(t) = 0$ for $t < 0$. Under this assumption, the convolution integral (\ref{eq:response_conv}) can be recast to
\begin{equation}
\label{eq:response_causal}
    \zeta(t) = \int_0^{\infty} f(\tau) \chi(t - \tau) \mathnormal{d}\tau = \int_{-\infty}^t f(t - \tau) \chi(\tau) \mathnormal{d}\tau.
\end{equation}
Due to causality and taking into account eq. (\ref{eq:ifft_ir_1}), the impulse response can be determined by using either only real or imaginary part of $\tilde{f}(\omega)$:
\begin{eqnarray}
\label{eq:ifft_ir_2}
f(t) &=& \frac{2}{\pi} \int_0^{\infty} f_R(\omega)\cos{(\omega t)} \textnormal{d}\omega \label{eq:cos_transform} \\ &=& - \frac{2}{\pi} \int_0^{\infty} f_I(\omega) \sin{(\omega t)} \textnormal{d}\omega. \label{eq:sin_transform}
\end{eqnarray}

\end{enumerate}

Note that for the sake of clarity, dependence on spatial variables and electrical conductivity pertinent to our application was omitted from the equations above.

In practice, we observed that using sine transform (\ref{eq:sin_transform}) results in a slightly better accuracy compared to the cosine transform (\ref{eq:cos_transform}) given the same filter length.

\section{Digital Linear Filters}
\label{sec:dlf}

In order to carry out the sine transform (\ref{eq:sin_transform}) efficiently, we applied the linear digital filter (DLF) method. DLF was introduced to geophysics by Ghosh in the early 70s \cite{ghosh1971dlfa, ghosh1971dlfb}, as a means of fast computations for geoelectric resistivity responses. The method was subsequently improved and expanded to other methods by many authors, and a lot of filters have been published. There are two particular developments, out of all these improvements, which are relevant for our application: (1) The kernel under consideration were early on always Bessel functions of some sort, and it was \citeA{anderson1973sinecosine} who first applied it to Fourier sine and cosine transforms. (2) If the kernel computation is very expensive the lagged-convolution type DLF introduced by \citeA{anderson1975improved} is very powerful, as additional times come at no or very little extra cost due to the reuse of the already computed kernels for new times. Although the use of DLF in geophysics is focused on Hankel and Fourier transforms in electromagnetics, the method itself works for any linear transform. 

\citeA{werthmuller2019tool} presented a tool to design filters for any linear transform provided that there exist (a) an analytical transform pair, or (b) a numerical computation in both domains with sufficient accuracy and precision over a wide range of argument values. We refer to that publication for an in-depth review of DLF in geophysics.

Using substitutions $t=e^x$ and $\omega=e^{-y}$ and multiplying by $e^x$ we can rewrite (\ref{eq:sin_transform}) as a convolution integral and approximate it by a $N$-point digital filter $\eta$ as \cite{anderson1975improved}
\begin{equation}
\label{eq:dlf}
    f(t) \approx -\frac{2}{\pi}\sum_{n = 1}^{N} \frac{f_I(b_n / t)\eta_n}{t},
\end{equation}
where the log-spaced filter abscissa values $b_n$ are a function of spacing $\Delta$ and shift $\nu$,
\begin{equation}
\label{eq:dlf_abscissa}
    b_n(\Delta, \nu) = \exp\left[ \Delta(- \lfloor (N + 1) / 2 \rfloor + n) + \nu \right].
\end{equation}

\begin{figure}[htbp]
  \centering
  \includegraphics[width=1.0\textwidth]{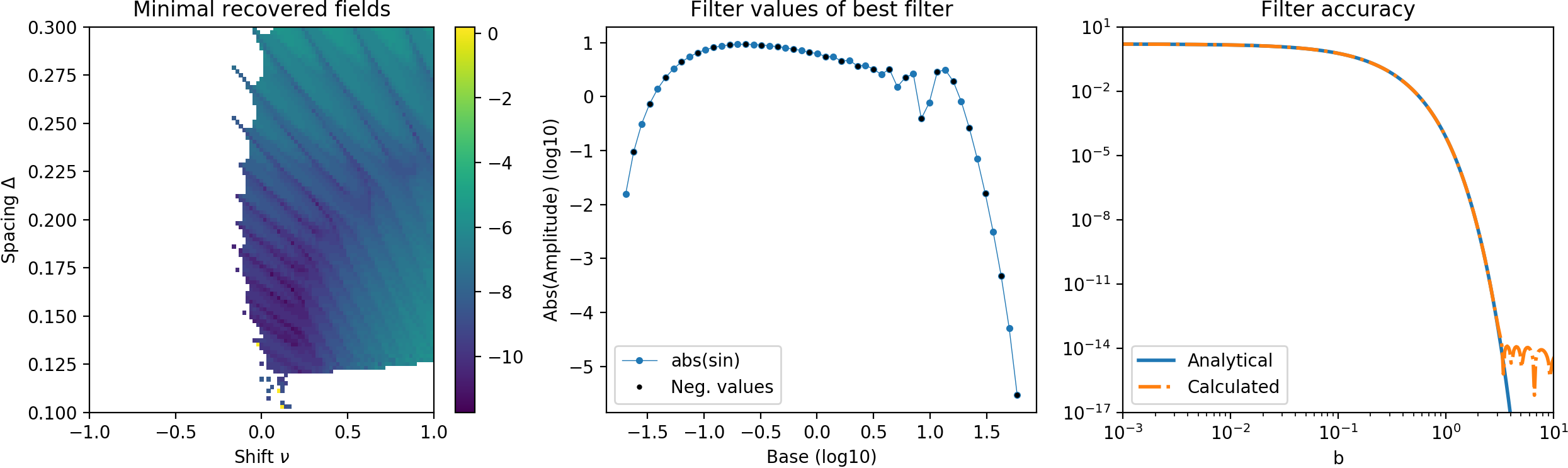} 
  \caption{Left: minimum recovered value of the analytic pair (\ref{eq:anpair}) as a function of spacing and shift. Center: filter values for the best filter with $\Delta = 0.114$ and $\nu = 1.07$. Right: the performance of the filter on the eq. (\ref{eq:anpair}).}
  \label{fig:filter}
\end{figure}

The optimal values for $\eta_n, \Delta$ and $\nu$ in eqs. (\ref{eq:dlf})-(\ref{eq:dlf_abscissa}) were found by following the method of \citeA{werthmuller2019tool}. In this work, we designed a 50-point filter such that it requires as few values of $\tilde{f}(\omega)$ as possible without compromising accuracy. To this end, we used the following analytic transform pair
\begin{equation}
\label{eq:anpair}
    \frac{\pi \exp\left(-ab\right)}{2} = \int^\infty_0 \frac{x}{a^2+x^2} \sin(xb) \mathnormal{d}x.
\end{equation}
The Figure \ref{fig:filter} shows the designed filter and its performance for the chosen analytic pair. 

Note that the \textit{naive} application of eq. (\ref{eq:dlf}) will require calculating $\tilde{f}(\omega)$ at  $N \times N_t$ frequencies, where $N_t$ is the length of an impulse response in time domain. This number can be drastically reduced by invoking the aforementioned lagged convolution approach. To give an example, for the one year long impulse response with the hourly time step (i.e., $N_t = 8766$) our filter required evaluating a maximum of 112 frequencies that range $\approx 12$ decades.

\begin{table}
\caption{Filter used in this study to approximate the sine transform \ref{eq:sin_transform}.}
\label{tab:sin_filter}
\begin{tabular}{|c|c|}
base & value \\ \hline
0.020351539057584585 & 0.015612515803531853    \\
0.02394618633584547  & -0.09462903411749914    \\
0.028175748203049585 & 0.3117770493488479      \\
0.0331523682171173   & -0.7331733032969688     \\
0.039007997604279074 & 1.3887195607048695      \\
0.045897893843668    & -2.256657066578928      \\
0.05400473719916161  & 3.2805358935454616      \\
0.06354347434512761  & -4.3780014125702555     \\
0.0747670175110597   & 5.472633981207025       \\
0.08797295025351011  & -6.487680055307104      \\
0.10351141765367007  & 7.379203081547922       \\
0.121794410143077    & -8.101406699379584      \\
0.1433066871108987   & 8.655711019132568       \\
0.16861862992378368  & -9.017610106427929      \\
0.19840136514614606  & 9.229490585407783       \\
0.23344455894136176  & -9.263211469329843      \\
0.2746773544586444   & 9.201625018217383       \\
0.3231930073442032   & -8.981017420159862      \\
0.38027787256818374  & 8.741193586707128       \\
0.44744551113066483  & -8.339636911178598      \\
0.5264768209596123   & 8.017252694339163       \\
0.6194672560404727   & -7.484385258748044      \\
0.728882385756068    & 7.168939484640828       \\
0.8576232675496689   & -6.5205283437188415     \\
1.009103366216787    & 6.298382148011787       \\
1.1873390592811512   & -5.517835729490835      \\
1.3970561281348315   & 5.467376261589812       \\
1.6438150584726328   & -4.53662000270438       \\
1.9341584722647591   & 4.6731213626935695      \\
2.2757846003123405   & -3.691290054788441      \\
2.6777513948763136   & 3.782371098562427       \\
3.1507166942679676   & -3.2281439933834775     \\
3.7072208071793002   & 2.5592546437619657      \\
4.36201900925792     & -3.232119665813557      \\
5.1324727678156865   & 1.5126346621525144      \\
6.039010067691346    & -2.2490157798136283     \\
7.105667043450954    & 2.624917369559576       \\
8.360725278884502    & -0.38778344764051975    \\
9.837461671301405    & 0.7697565678585556      \\
11.575030742696057   & -2.8607107883073972     \\
13.619502791580832   & 3.1199881060456427      \\
16.025085411278443   & -1.9120353599090574     \\
18.855560762285485   & 0.809753326019789       \\
22.185976706870886   & -0.2641923369068591     \\
26.104636645033704   & 0.07088122130542221     \\
30.715440810780017   & -0.016121403161804555   \\
36.14064110637662    & 0.0030871244528753875   \\
42.524082516878885   & -0.00047298225758749283 \\
50.03501704852909    & 5.153158194873856e-05   \\
58.87259131464845    & -2.9723596837103312e-06
\end{tabular}
\end{table}

%

%
%

\acknowledgments
This work was supported by the ESA through the Swarm DISC project. The staff of the geomagnetic observatories and INTERMAGNET are thanked for supplying high-quality observatory data, and BGS are thanked for making quality-controlled observatory hourly mean values openly available. Satellite and observatory input data used in this study as well as estimated time series of SH coefficients can be retrieved from \url{https://doi.org/10.5281/zenodo.4047833}. Subsurface conductivity model is available at \url{https://doi.org/10.5281/zenodo.4058852}. We thank Chris Finlay, Nils Olsen and Jakub Vel{\'i}msk{\'y} for insightful discussions that facilitated this study.

%
%


\end{document}